\documentclass[aps,pra,10pt,twocolumn,noeprint]{revtex4-1}
\usepackage{amsmath,amssymb,amsfonts,graphicx,hyperref,times}
\usepackage{amsmath,amssymb,amsfonts,bbm,graphicx,hyperref,times}
\usepackage[utf8]{inputenc}
\usepackage[T1]{fontenc}
\usepackage{microtype}
\usepackage{bm}
\usepackage{xcolor}

\newcommand{\id}{\mathbbm{1}}

\newcommand{\Tr}{\mathrm{Tr}}
\renewcommand{\vec}{\boldsymbol}
\let\originalleft\left
\let\originalright\right
\renewcommand{\left}{\mathopen{}\mathclose\bgroup\originalleft}
\renewcommand{\right}{\aftergroup\egroup\originalright}
\renewcommand{\right}{\aftergroup\egroup\originalright}

\newtheorem{definition}{Definition}

\begin{document}
\title{Resource theory of quantum non-Gaussianity and Wigner negativity}
\author{Francesco Albarelli}\email{francesco.albarelli@unimi.it}
\affiliation{Quantum Technology Lab, Dipartimento di Fisica {``Aldo Pontremoli''}, Universit\`a degli Studi di
 Milano, I-20133 Milano, Italy}
\author{Marco G. Genoni}\email{marco.genoni@fisica.unimi.it}
\affiliation{Quantum Technology Lab, Dipartimento di Fisica {``Aldo Pontremoli''}, Universit\`a degli Studi di
 Milano, I-20133 Milano, Italy}
\author{Matteo G. A. Paris}\email{matteo.paris@fisica.unimi.it}
\affiliation{Quantum Technology Lab, Dipartimento di Fisica {``Aldo Pontremoli''}, Universit\`a degli Studi di
 Milano, I-20133 Milano, Italy}
\author{Alessandro Ferraro}\email{a.ferraro@qub.ac.uk}
\affiliation{
School of Mathematics and Physics, Queen's University, 
BT71NN Belfast, U.K.}
\begin{abstract}
We develop a resource theory for continuous-variable systems grounded on operations routinely available within current quantum technologies. In particular, the set of free operations is convex and includes quadratic transformations and conditional coarse-grained  measurements. The present theory lends itself to quantify both quantum non-Gaussianity and Wigner negativity as resources, depending on the choice of the free-state set --- \textit{i.e.}, the convex hull of Gaussian states or the states with positive Wigner function, respectively. After showing that the theory admits no maximally resourceful state, we define a computable resource monotone --- the Wigner logarithmic negativity. We use the latter to assess the resource 
content of experimentally relevant states --- \textit{e.g.}, photon-added, photon-subtracted, cubic-phase, and cat states --- and to find optimal working points of some resource concentration protocols. We envisage applications of this framework to sub-universal and universal quantum information processing 
over continuous variables.
\end{abstract}
\maketitle
\section{Introduction}
Gaussian states of bosonic systems have played a pivotal role for 
continuous-variable (CV) quantum technology since its inception.
The theoretical analysis of Gaussian states 
is made simpler by the fact that they can be compactly described by first and second statistical moments, yet they can manifest many genuinely quantum properties ~\cite{Ferraro2005,Weedbrook:12,adesso2014continuous, Genoni2016, Robinson1965, Holevo_new_book}. Experimentally, the generation and manipulation of Gaussian states has been made possible by the availability of second order non-linearities (Gaussian-preserving operations) in various physical platforms, including optical, atomic, and opto-mechanical systems \cite{serafini2017quantum}. In particular, the unconditional character of these operations has enabled the realization of a variety of quantum-information protocols \cite{Braunstein:05,Weedbrook:12}.
\par
Notwithstanding the rich phenomenology observed with Gaussian states 
and operations, the restriction to this setting entails several no-go theorems for relevant tasks, including entanglement distillation~\cite{Eisert2002,Fiurasek2002,Giedke2002}, error correction~\cite{Niset2009}, computation~\cite{Lloyd1999,Menicucci2006}, and few others~\cite{Adesso2009b,Magnin2010,Jabbour2014,lami2018gaussian}.
Moreover, several CV protocols improve their performances when used
in conjunction with non-Gaussian states and operations. These include estimation problems~\cite{Adesso2009,Genoni2009,Rossi2016}, teleportation~\cite{Opatrny2000,Cochrane2002,Olivares2003} and cloning~\cite{Braunstein2001,Cochrane2004} as well as more foundational schemes 
as those aimed at testing Bell and Leggett-Garg inequalities~\cite{Munro1999,Wenger2003,Acin2009,Paternostro2009}.
\par
For all the above reasons, non-Gaussian states and operations are 
considered crucial resources for the development of CV quantum 
information technologies, leading to major research efforts towards
their understanding, characterisation, and experimental generation. 
In particular, quantum optical schemes as photon-subtraction 
and photon-addition have received attention~\cite{Olivares2003,Kim2008,Ourjoumtsev2006,NeergaardNielsen2006,Ourjoumtsev2007a,Sabapathy2017}. Different ways to quantify non-Gaussianity~\cite{Genoni2007,Genoni2008,Genoni2010,Marian2013a,Ivan2012} have been introduced, 
and exploited to assess the properties of experimentally generated non-Gaussian states~\cite{Allevi2010,Allevi2010b,Barbieri2010}.
\par
However, for reasons to be detailed below, no satisfactory resource theory
of non-Gaussianity has been developed yet. For example, questions such as ``Given two non-Gaussian states, which one is more resourceful?'' and ``Is 
it possible to transform a given non-Gaussian state to another one with Gaussian transformations, and at which rate?'' remain unanswered. This represents a major obstacle to our understanding of non-Gaussian resources and the development of applications thereof. The present work aims at overcoming this obstacle by introducing a general theoretical framework 
for non-Gaussian resources.
\par
Resource theories~\cite{Coecke2014a} are a powerful framework to study manipulation of quantum states under some operational restriction on the allowed operations. Entanglement theory is the prototypical example but many others have been recently developed. In particular, general results have been obtained for a vast class of resource theories both in the asymptotic~\cite{Brandao2015b} and single-shot regime~\cite{Gour2017}. In general, a resource theory stems from two interlinked starting points. First, the identification of a set of operations that are regarded, for reasons that depend on the setting at hand, as readily available (\textit{free operations}). For example, these are local operations and classical communication in the resource theory of entanglement. Second, the classification of all possible states in two categories: \textit{free states}, that are considered freely available (typically via free operations) and non-resourceful, and \textit{resource states}. Separable and entangled states are an example of this classifications. The quantification and manipulation of resources via free operations are the central concerns of resource theories.
\par
As a matter of fact, there are at least two major difficulties towards such a theory in the case of non-Gaussian resources. First, it is natural to identify the set of Gaussian states as free states. However, this set is not convex and therefore non-Gaussianity \textit{per se} cannot be considered a quantum resource of practical relevance in general: in fact, non-Gaussianity can be generated by classical randomness, which is a 
readily available in most contexts and therefore, from an operational 
view-point, free.
Second, the intrinsic infinite-dimensional character of CV systems implies that some operations (conditioning on the continuous real outcomes of measurements, see below) that are ideally free in various resource theories are, in this context, unfeasible. These two roadblocks have hindered the development of a satisfactory resource theory, namely a theory that is both general and of practical relevance in realistic settings. Here we overcome these roadblocks by adopting a pragmatic approach and incorporating them in our definitions of free operations and states.
\par
Our analysis is based on tools developed to describe quantum systems via their phase-space representation. There, quasiprobability distributions play a central role and their non-positivity is considered as a characteristic trait of quantum theory~\cite{Ferrie2008,Ferrie2011a,DeBrota2017}, directly linked to its contextual character~\cite{Spekkens2007,Delfosse2016,Howard2014}. In particular, the connection between negativity of the Wigner function and Gaussian states is strong: pure states with a positive Wigner function are only Gaussian, as stated by the Hudson theorem~\cite{Hudson1974,Soto1983}.
However, complications arise when dealing with mixed states, since there is no Hudson theorem for mixed states~\cite{Brocker1995a,Mandilara2009,Mandilara2010}.
In particular there exist mixed states that are not mixtures of Gaussian states, yet have a positive Wigner function~\cite{Filip2011,Jezek2011,Genoni2013,Hughes2014,Palma2014,Happ2018}. Despite this, we are going to introduce a computable resource quantifier based on the negativity of the Wigner function and we are going to show that it is a proper monotone for the resource theory at hand.
\par
Our framework is based on an analogous one that has already revealed its efficacy in the context of finite-dimensional systems (discrete variables, DVs). In particular, the formidable task of identifying the resource responsible for quantum advantage in DV quantum computation has led to the formulation of various resource theories~\cite{Veitch2014,Howard2017,Ahmadi2017,Stahlke2014}. 
In this context, stabilizer states and Clifford unitaries play the role of Gaussian states and unitaries, respectively.
On the other hand, so-called \textit{magic states} play the role of non-Gaussian states.
For finite dimensional systems it is natural to define Wigner functions on a finite dimensional state space~\cite{Gibbons2004,Miquel2002} (opposed to the CV phase space); in particular for odd-dimensional systems a DV version of the Hudson theorem was proven~\cite{Gross2006}.
For such systems, it has been shown that the negativity of the discrete Wigner function is necessary to obtain a circuit which is both universal and cannot be efficiently simulated with known classical algorithms~\cite{Veitch2012a,Mari2012,Pashayan2015}.
Based on this, a computable monotone ---dubbed \textit{Wigner logarithmic negativity} (WLN)--- has been identified~\cite{Veitch2014}.
The monotone that we introduce here can be considered as a CV counterpart of an analogous quantity for DV known as ``mana''. 
\par
After developing the general framework, we apply it to two tasks.
First, we assess and compare the resourcefulness of various non-Gaussian states that have been theoretically proposed or even realized experimentally.
In particular, we are going to see that the cubic phase state ---a resource that unlocks universality in the context of CV measurement-based quantum computation--- has a degree of resourcefulness that can be non-trivially boosted by squeezing operations.
Also, we give evidences that photon-added and photon-subtracted states are at most as resourceful as a single-photon Fock state.
Second, concerning state manipulation, we evaluate the efficiency of a set of Gaussian protocols that consume copies of non-Gaussian inputs to produce more resourceful outputs. These types of resource-concentration protocols have been shown to be crucial in the context of other resource theories, such as for quantum communication and fault-tolerant DV quantum computation. Our results individuates optimal working points for these concentration protocols, and can thus be used to guide the development of new more efficient ones.
\par
The paper is structured as follows. In Section \ref{s:rt} we recall the
main ingredients to build a resource theory: free states, free operations 
and resource monotones, and apply them to the development of a
resource theory of quantum non-Gaussianity and Wigner negativity. In particular, we show the absence of maximally resourceful states and  
introduce a computable monotone, the {\em Wigner logarithmic negativity}. Section \ref{s:ps} is devoted to resource analysis of some classes of relevant pure states, including cubic phase-state, photon added/subtracted Gaussian states and
cat states. We also devote attention to compare the corresponding value
of monotones at fixed energy. In Section \ref{sec:concentration} we focus
on concentration protocols based on passive Gaussian operations in order
to assess their performances in the task of negativity concentration.
Finally, Section \ref{s:out} closes 
the paper with some concluding remarks.
Generalities about Gaussian states and phase space formalism, a discussion of our results in comparison with other resource theories and technical results about the monotones are contained in three appendices.
\section{Resource theoretical framework}\label{s:rt}
As said, a resource theory is composed of three main elements: free operations, free states, and resources.
The fundamental condition that links these elements is that the set of free states must be closed under the set of free operations ---\textit{i.e.}, it must not be possible to obtain a resource state by applying only free operations.
This restriction still leaves much freedom in the choice of both free operations and states, and the structure of the resource theory is dictated by this choice. For example, free states can be chosen as the maximal set of states generated by free operations, or vice-versa free operations can be considered as the maximal set that leaves the free states invariant. The freedom left by the abstract formulation is generally restricted by operational issues. Crucially, we are going to be guided by practical considerations regarding the prompt availability of operations such as classical randomness and conditional measurements.

In this section, we introduce a resource theory based on two pragmatical considerations: first, we consider operationally relevant Gaussian transformations, including in particular those that involve conditioning on coarse-grained measurements; second, we consider convex free-state sets.
These features set apart our framework from previous studies on the quantification and manipulation of non-Gaussian resources \cite{Genoni2007,Genoni2008,Genoni2010}.

\subsection{Free operations}
The set of free operations is our starting point for the construction of the resource theory. In general, there is a contrast between physically motivated sets of operations and other larger sets with better mathematical properties. 

Physically motivated operations are those which, in a given context, can be assumed to be implementable without effort ---for example, SLOCC operations in entanglement theory, or stabilizer operations in the resource theory of stabilizer computation. 
Unfortunately, these operations are usually hard to characterize and other sets of free operations, with better mathematical properties, are often introduced (maximal resource non-generating operations). Even though they do not generate any resource, some amount of resource is typically needed to practically implement them.
As said, here we adopt an operational point of view, hence we are not going to deal with these maximal resource non-generating operations.
\par
Standard notation for bosonic systems is used throughout, and a brief review of the quantum phase space formalism is given in the Appendix~\ref{app:phase_space}, with emphasis on the Gaussian sector.
We denote with $| \psi_G \rangle$ an arbitrary pure Gaussian state, with $U_G$ a Gaussian unitary, and with $\mathcal{S}\left( \mathcal{H} \right)$ the set of density operators on the Hilbert space $\mathcal{H}$ of an arbitrary (finite) number of bosonic modes.
In the context of quantum optics a set of free operations with a strong operational motivation is given by \textit{Gaussian protocols} (GPs), which we define as follows.
\begin{definition}
\label{def:GP}
A Gaussian protocol is any map from $\rho \in \mathcal{S} \left( \mathcal{H} \right)$ to $\sigma \in \mathcal{S} \left( \mathcal{H}' \right)$ composed of the following operations:
\begin{enumerate}
	\item Gaussian unitaries: $\rho \to U_G\,\rho \,U_G^\dagger$.
	\item Composition with pure Gaussian states: $\rho \to \rho \otimes | \psi_G \rangle \langle \psi_G | $.
    \item Pure Gaussian measurements on subsystems: $ \rho \to \Tr_S \left[ \rho \, \id \otimes \left| \psi_G \left( \boldsymbol{\alpha }\right) \rangle \langle \psi_G \left( \boldsymbol{\alpha }\right) \right| \right] / p(\boldsymbol{\alpha} | \rho)$, with probability density $ p(\boldsymbol{\alpha} | \rho )=\Tr \left[ \rho \, \id \otimes \left| \psi_G \left( \boldsymbol{\alpha }\right) \rangle \langle \psi_G \left( \boldsymbol{\alpha }\right) \right| \right]$ where $\boldsymbol{\alpha }$ is a vector of continuous measurement outcomes in the real domain.
    \item Partial trace on subsystems: $\rho \to \Tr_S \left[ \rho \right]$.
    \item The above quantum operations conditioned on classical randomness or 
    	\begin{enumerate}
    		\item single measurement outcomes (ideal case)
    		\item measurement outcomes falling into finite-size intervals (operational case).
    	\end{enumerate}
\end{enumerate}
\end{definition}

The above operations encompass what is routinely available in current experiments where CVs are manipulated at the quantum level.
They include general quadratic interactions, the generation and control of a large number of bosonic systems, and efficient measurement strategies such as homodyne detection, which correspond to a projection on an infinitely squeezed and unnormalized Gaussian state $| \psi_G \rangle$.
Actually, requirement \textit{3.} could be restricted to homodyne detection only, since the projection on any Gaussian state can be obtained via Gaussian unitaries and homodyne detection~\cite{Giedke2002}.
In particular, in quantum optical setups, such Gaussian unitaries correspond to inline squeezing operations and passive linear optics circuits \cite{Braunstein2005}.
Notice that probabilistic operations are included, therefore a generic \textit{probabilistic} GP $\Lambda_\text{GP}$ is a trace non-increasing CP map.
Physically, nondeterministic maps come from selecting particular states based on measurement outcomes; this operation cannot generate resource states from free states, not even probabilistically.
However we are going to see that it is possible to use nondeterministic free operations to probabilistically concentrate the resource.

The requirements $\textit{1.-4.}$ above are standard in the context of non-Gaussianity quantification \cite{Genoni2007,Genoni2008,Genoni2010}, whereas requirement $\textit{5.}$ needs an explanation. 
First and foremost, a reasonable request for a \emph{quantum} resource theory is that \emph{classical} randomness should not be regarded as a resource, therefore the inclusion of conditioning on classical randomness. In turn, as anticipated, the latter entails that set of GPs is \textit{convex}. Second, GPs are the CV analogous of the stabilizer operations introduced in Ref.~\cite{Veitch2014} but with some relevant differences due to the infinite dimensional setting. Since the outcomes of projective Gaussian measurements are continuous parameters, single outcomes are obtained with zero probability.
Therefore the class of GPs satisfying property $\textit{5.a}$ (\textit{ideal} GPs) contains operations that are unattainable practically; for this reason, in the applications of our resource theory, we mainly consider the subclass of GPs that satisfy condition $\textit{5.b}$ ---which we dub \emph{operational} GPs.
This subclass is defined by the requirement that every output state must be obtained with finite probability, therefore every conditioning must be done on a finite-size interval of measurement outcomes. The major consequence of choosing operational GPs is that they exclude the possibility to define a resource theory on pure states only, since output states with nonzero probability must be mixed. We remark that this is a peculiar feature of the framework here introduced that is inherently linked to infinite-dimensional systems, without analogue in the resource theories introduced so far in DV settings.

To conclude this part, we stress again that, from an operational point of view, it is not useful to enlarge the set of GPs, since  any operation ``easy'' to implement in the lab is already included. Not surprisingly however, these physically meaningful free operations are hard to characterize mathematically ---similarly to what occurs for other resource theories. A more detailed discussion on the maximal sets and a comparison with other resource theories is left to Appendix \ref{app:maximalset}.

\subsection{Free states}

We consider two classes of free states, both satisfying the standard requirements \cite{Brandao2015b} ---in particular, both being closed with respect to GPs. The first class is the most natural choice, and it is given by the maximal set of states that can be generated via GPs ---namely, the Gaussian \textit{convex} hull, defined as
\begin{equation}
\label{eq:Gconvhull}
\mathcal{G} = \left\{ \rho \in \mathcal{S}\left( \mathcal{H} \right) \, | \, \rho = \int \! \mathrm{d} \boldsymbol{\lambda} \,  p \left(\boldsymbol{\lambda} \right) \left| \psi_G \left( \boldsymbol{\lambda} \right) \rangle \langle \psi_G \left( \boldsymbol{\lambda} \right) \right| \right\},
\end{equation}
where $p\left( \boldsymbol{\lambda} \right) $ is an arbitrary probability distribution.
We remark that $\boldsymbol{\lambda}$ in Eq.~\eqref{eq:Gconvhull} represents the vector of $2n^2 + 3n$ real parameters needed to parametrize an arbitrary $n$-mode pure Gaussian state.
We dub continuous variable quantum states not in the Gaussian convex hull $\mathcal{G}$ as \textit{quantum non-Gaussian} (QnG). Therefore, the theoretical framework derived considering set $\mathcal{G}$ will be referred to as the \textit{resource theory of quantum-non-Gaussianity}. We recall that witnesses of QnG states have previously been introduced~\cite{Filip2011,Jezek2011,Genoni2013,Hughes2014,Palma2014,Happ2018}, albeit outside of a resource theoretical context. 

Alternatively, one can define the free states as those with a positive Wigner function:
\begin{equation}
\label{eq:poswig_set}
	\mathcal{W}_+ = \left\{ \rho \in \mathcal{S} \left( \mathcal{H} \right) \, | \, W_\rho \left( \boldsymbol{r} \right) \geq 0  \right\},
\end{equation}
where $W_\rho$ is the Wigner function of the state $\rho$; this set is also convex and it is a proper superset of $\mathcal{G}$, i.e. $\mathcal{G} \subset \mathcal{W}_+ $.
As per the Hudson theorem~\cite{Hudson1974,Soto1983}, these two sets coincide when restricted to pure states.
We dub continuous variable quantum states with a non-positive Wigner function (states not in the set $\mathcal{W}_+$) as \textit{Wigner negative} (WN). Therefore, the theoretical framework derived considering $\mathcal{W}_+$ will be referred to as the \textit{resource theory of Wigner negativity}. 

We recall that, from an operational point of view, a positive Wigner function is a sufficient condition to have a quantum system that can be efficiently simulated by classical algorithms~\cite{Mari2012,Veitch2013,Pashayan2015,Rahimi-Keshari2015}.
As already mentioned, not all Wigner-positive states can be generated using GPs.
In this sense, the choice of $\mathcal{W}_+$ as free states of the theory is less natural with respect to choosing $\mathcal{G}$.
However this is still a valid choice, in the same sense that, for the entanglement resource theory, it would be a valid choice to consider the set composed of states with positive partial transpose.
QnG states with positive Wigner function are \textit{bounded} resources: despite the fact that they cannot be generated using GPs, no other resource (\textit{free} resource) can be extracted from them using GPs only ---not even when an arbitrary large supply of them is available.

The definition of the set $\mathcal{G}$ allows to specify some additional considerations regarding GPs.
An arbitrary GP can be extended to a \textit{deterministic} (trace-preserving) GP $\Lambda_\text{DGP}$ by considering all the possible outcomes.
A deterministic GP can be characterized in terms of its free Kraus operators; free means that the trace-decreasing map given by a single Kraus operator is a GP itself.
By considering Kraus operators corresponding to all the continuous outcomes of Gaussian measurements, we can write $\Lambda_\text{DGP} \left( \rho \right) = \int \mathrm{d} \vec{\lambda} K_{\vec{\lambda}}  \rho K_{\vec{\lambda}}^\dag$, where $K_{\vec{\lambda}} \mathcal{G} K_{\vec{\lambda}}^\dag \subset \mathcal{G}$ (disregarding normalization).
A different free Kraus representation of $\Lambda_\text{DGP}$ can be obtained by coarse-graining measurement outcomes, $\Lambda_\text{DGP} \left( \rho \right) = \sum_i  K_i \rho K_i^\dag$, with where $K_i\mathcal{G} K_i^\dag \subset \mathcal{G}$; in this case every Kraus operator corresponds to an operational GP, \textit{i.e.} it gives an output state with finite probability.

\subsection{Absence of maximally resourceful states}

A relevant observation that can be made at this stage is that no maximally resourceful states exist in the present resource theory ---a result that is at odds with the most common resource theories, including entanglement. Namely, there is no resource state that can be transformed via GPs into any other state, in particular any other pure state.

For operational GPs, this is an immediate consequence of the fact that the output states are either mixed (when measurements are involved) or they are the output of a Gaussian unitary operation on the given state. In the latter case a parameter-counting argument immediately proves the claim since, on one hand, Gaussian unitaries on a finite number $n$ of bosonic systems are characterised by a finite number of parameters (namely, the dimension of the affine symplectic group ${\rm ISp}(2n,\mathbb{R})$: $2n^2+3n$), whereas, on the other hand, a generic pure CV state cannot be specified by a finite number of parameters, due to the infinite dimension of the Hilbert space.
A slightly refined argument is valid also for the case of ideal GPs. Again, it is sufficient to consider the case of pure output states. In fact, any ideal GP with pure outputs is an element of the set of (non necessarily positive) linear bounded superoperators $\Phi$ that send the set of Gaussian states into itself. These maps $\Phi$ have been studied in details in Ref.~\cite{DePalma2015c}, where it is proven that they are characterised by a finite number of parameters. Therefore, again a parameter-counting argument proves the claim.

The absence of a maximally resourceful state implies relevant consequences that are peculiar to the present resource theory. First, regarding resource quantification, there exist no natural unit for the resource at hand to which all measures can be normalized. Second, regarding resource manipulation, there exist no natural target for resource distillation protocols, nor a natural starting state for resource dilution.

Despite this, notice that there exist at least one class of states that can play the role of maximally resourceful states, the so-called cubic-phase states~\cite{Gottesman2001} (see below). It is in fact known that, provided that an arbitrary large supply of these states can be consumed, any state can be generated via ideal GPs~\cite{Menicucci2006} (as recalled, magic states play an analogous role for DV stabilizer protocols). In Sec.~\ref{ssec:cps} we are going to consider cubic-phase states more closely. A similar result is suggested to hold true also for Fock states \cite{Ghose2006}. 

\subsection{Monotones}\label{subsec:monotones}
Once the sets of free operations and states are chosen, one can indeed try to quantify a resource.
In general, there is no unique way to quantify a resource and different monotones are connected to the performance of different tasks.
Moreover, monotones can be used as a tool to assess the feasibility of resource conversion.
In the best case scenario, a complete set of monotones can give necessary and sufficient conditions for the conversion between resource states, as in~\cite{Ahmadi2017}.

We can now define both Quantum-non-Gaussianity and Wigner Negativity monotones. 
\begin{definition}\label{def:monotone} A Quantum-non-Gaussianity (resp. Wigner Negativity) monotone is a functional from the set of quantum states to non-negative real numbers $\mathcal{M} : \mathcal{S}\left( \mathcal{H} \right) \to [0,\infty)$ which satisfies the following properties
\begin{enumerate} 
	\item $\mathcal{M} ( \rho ) =0 \quad \forall \rho \in \mathcal{G}$ (resp. $\mathcal{W}_+$).
	\item (Monotonicity under deterministic Gaussian protocols) 
	\\For any trace-preserving GP $\Lambda_\text{DGP}$ the monotone must not increase: $\mathcal{M} \left( \rho \right) \geq 	\mathcal{M}\left( \Lambda_\text{DGP} \left(  \rho \right) \right)$.
	\item (Monotonicity on average under probabilistic Gaussian protocols)\\
	Given a trace-preserving GP $\Lambda_\text{DGP}$ we can express its action in terms of free Kraus operators, we require that the monotone must not increase on average:
	\begin{enumerate}
		\item Ideal case: $\Lambda_\text{DGP} \left( \rho \right) = \int \mathrm{d} \vec{\lambda} p (\vec{\lambda} | \rho ) \sigma_{\vec{\lambda}} $, where $\sigma_{\vec{\lambda}} = \frac{1}{p({\vec{\lambda}}|\rho)} K_{\vec{\lambda}}  \rho K_{\vec{\lambda}}^\dag$. 
		We require that 
	 $\mathcal{M} \left( \rho \right) \geq \int \mathrm{d} \vec{\lambda} p (\vec{\lambda} | \rho )  \mathcal{M}\left( 
		\sigma_{\vec{\lambda}} \right)$.
			\item Operational case: $\Lambda_\text{DGP} \left( \rho \right) = \sum_i  p_{i | \rho} \sigma_{i}$, where $\sigma_i = \frac{1}{p_{i | \rho}} K_i   \rho K_i^\dag$.
			We require that 	 $\mathcal{M} \left( \rho \right) \geq \sum_i p_{i | \rho} \mathcal{M}\left( 
		\sigma_i \right)$
		\end{enumerate}
\end{enumerate}
\end{definition}

Some additional properties that a monotone can enjoy are \emph{faithfulness}: $\mathcal{M}\left( \rho \right) > 0 \iff \rho \notin \mathcal{G}$ (\textit{resp.} $\mathcal{W}_+$), \emph{convexity}: $\mathcal{M} \left(  \int \mathrm{d} \nu p(\nu) \rho_{\nu} \right) \leq \int \mathrm{d} \nu p(\nu) \mathcal{M} \left( \rho_{\nu} \right)$ for a generic probability distribution $p(\nu)$ and \emph{additivity}: $\mathcal{M} \left( \rho \otimes \sigma \right) = \mathcal{M} \left( \rho \right) + \mathcal{M} \left( \sigma \right) $\footnote{In the context of entanglement theory this property is called strong additivity, while weak additivity only requires $\mathcal{M} \left( \rho^{\otimes n} \right) = n \mathcal{M} \left( \rho \right)$.}.
If the monotone is convex, monotonicity on average directly implies monotonicity under deterministic operations (\textit{3a} $\implies$ \textit{2}), moreover convexity also gives operational average inequalities from the ideal ones (\textit{3a} $\implies$ \textit{3b}).

Monotones can be used to give bounds on the efficiency of interconversion between resource states.
Suppose that $\Lambda$ is a free operation which converts resource states in a probabilistic manner: it maps $k$ copies of $\rho$ to $m$ copies of a target state $\sigma$, i.e. $\Lambda \left( \rho^{\otimes k} \right) = \sigma^{\otimes m}$ with probability $p$.
By virtue of the monotonicity on average (\textit{3b}) we can write
\begin{equation}
\label{eq:eq:resbound1}
\mathcal{M} \left( \rho^{\otimes k} \right) \geq p \mathcal{M} \left( \sigma^{\otimes m} \right),
\end{equation}
where we considered an operational GP to get a finite probability $p$ and we discarded the other conditional states in the sum.

Moreover, additive monotones allow us to express the inequality in terms single letter quantities
\begin{equation}
\label{eq:eq:resbound_additive}
k \mathcal{M} (\rho) \geq p \, m \mathcal{M} ( \sigma).
\end{equation}
This inequality also gives a lower bound for the average conversion ratio~\cite{Veitch2014}.
On average we will need to run the probabilistic operation $1 / p$ times to obtain a successful outcome, therefore the average number $n$ of copies needed to extract $m$ target states is $\mathbb{E} \left[ n \right]= k / p$.
We can thus rewrite~\eqref{eq:eq:resbound_additive} as  
\begin{equation}
\label{eq:add_mon_bound}
\mathbb{E} \left[ n \right] = \frac{k}{p} \geq m \frac{\mathcal{M}\left( \sigma \right)}{\mathcal{M}\left(\rho\right)},
\end{equation}
\textit{i.e.}, the average number of copies of the input state is lower bounded by the ratio of the monotones times the number of output copies of the protocol.
This means that in order to concentrate the resource (\textit{i.e.}, $\mathcal{M}\left(\sigma\right) / \mathcal{M}\left(\rho\right) > 1$), we need an average conversion ratio smaller than unity $ m / \mathbb{E} \left[ n \right] < 1 $.

If free operations converting two resource states in both directions exist, we must have $\mathcal{M} (\rho^{\otimes k}) = \mathcal{M} (\sigma^{\otimes m})$; this is trivially true if the conversion is achieved with a free unitary transformation. It is usually difficult to exactly convert between resource states using a finite number of copies, therefore it is customary to consider conversions in the asymptotic limit of infinite copies.
However, we are not going to deal with the asymptotic resource theory of QnG in the present work.

\subsubsection{A computable monotone: Wigner logarithmic negativity}
Negativity of the Wigner function has long been recognized as an important quantum feature and in particular the volume of the negative part has been introduced as a nonclassicality quantifier~\cite{Kenfack2004}. Here we use it to define a resource monotone.

In~\cite{Veitch2014} a computable and additive magic monotone based on the negative values of the discrete Wigner function, dubbed \emph{mana}, was introduced.
We call the CV counterpart \emph{Wigner logarithmic negativity} (WLN); it is defined as \footnote{The base of the logarithms is irrelevant for this definition and the following properties; however, in the numerical results we choose logarithms in base 2.}
\begin{equation}
\label{eq:CV_mana}
\mathsf{W} \left ( \rho \right) = \log \left( \int \! \mathrm{d} \boldsymbol{r} \, \left| W_\rho \left( \boldsymbol{r} \right) \right| \right),
\end{equation}
where the integral runs over the whole phase-space $\mathbb{R}^{2n}$, where $n$ is the number of modes.
In Appendix~\ref{app:monotones} we show that this monotone satisfies all the required properties even if it is not convex.
The proofs rely on the fact that the negativity $\mathcal{N}[\rho] = \int \mathrm{d} \vec{r} | W_\rho (\vec{r}) | - 1 \,$ is also a monotone, which is convex but not additive.
The crucial monotonicity properties \textit{3.} for $\mathsf{W}$ then follow thanks to Jensen's inequality for the logarithm.

Clearly, the WLN is a faithful monotone for the resource theory of Wigner Negativity but not for Quantum-non-Gaussianity. This is akin to what happens in entanglement theory for the log-negativity of entanglement~\cite{Plenio2005a,Zhu2017a}, depending on whether one considers separable or positive partial transpose states as free states.

The WLN is an additive monotone, since the Wigner function of separable states can be factorized.
This means that the bound~\eqref{eq:add_mon_bound} is valid and we can use the ratio between logarithmic Wigner negativities to lower bound the average number of copies of an input resource state to obtain a certain number of copies of the target state using a probabilistic Gaussian protocol.
We remark that this result does not say anything about the actual existence of such protocols.

Similarly to the DV case we can prove that the WLN is essentially the unique measure which depends on the negative values of the Wigner function, under the assumption that the position of these ``negative patches'' in phase space do not affect such a measure.
The proof follows the same idea of the DV case presented in~\cite{Veitch2014} and lately extended to coherence and entanglement~\cite{Zhu2017a} (see Appendix~\ref{app:mana} for further details).

As a final remark, notice that the WLN is \emph{computable} in the sense that its value can usually be assessed by numerical integration.
However, in general it will be prohibitively hard to obtain closed-form expressions, since the analytical integration of the absolute value of a function is hindered by the requirement of finding the zeros of such function.

\subsubsection{Faithful Quantum non-Gaussianity monotones}
We want here to mention two possible ways to define \emph{faithful} QnG monotones.

The relative entropy of a state from the set of Gaussian states defines a proper measure of non-Gaussianity~\cite{Genoni2010}, which we call relative entropy of non-Gaussianity~\cite{Marian2013a}.
Note however that, since the set of states with a Gaussian Wigner function is not convex, the relative entropy of non-Gaussianity can be arbitrarily increased by GPs.
This measure is particularly simple for pure states:
\begin{equation}
\label{eq:nong}
\delta \left[ |\psi \rangle \right] = S\left( \rho || \tau_G \right) = S\left( \tau_G \right),
\end{equation}
where $S\left( \rho || \sigma \right)=  \mathrm{Tr} \left[ \rho \left( \log \rho - \log \sigma \right) \right]$ is the quantum relative entropy, $S(\rho) = \mathrm{Tr} ( \rho \log \rho)$ is the von Neumann entropy and $\tau_G$ is the reference (mixed) Gaussian state having the same covariance matrix as $| \psi \rangle$.
We refer the reader to Appendix~\ref{app:Gaussian} for a short introduction to Gaussian states and details on the von Neumman entropy of Gaussian states.
We believe it should be possible to build a QnG monotone by extending this measure to mixed states with a convex roof construction, i.e.
\begin{equation}
\label{eq:convroof}
\delta_{\text{CR}} \left [ \rho \right] = \inf_{ p_i , | \psi_i \rangle} \sum_i p_i \delta \left[ | \psi_i \rangle \right]
\end{equation}
where $\rho = \sum_i p_i |\psi_i \rangle \langle \psi_i|$; $i$ can also represent a continuous value, in which case $p_i$ becomes a distribution and the sum is replaced by an integral.
The functional $\delta_{\text{CR}}$  is convex by construction and property \textit{1} and \textit{2} of Definition~\ref{def:monotone} can easily be proven.
We have not been able to prove property \textit{3a} (property \textit{3b} follows by convexity), however we performed some preliminary numerical checks and we conjecture property \textit{3a} to be true (see Appendix~\ref{app:numcheck_convex} for more details).

We also mention that a different approach to introduce a faithful monotone could be to connect the resource theory of quantum non-Gaussianity to the resource theory of coherence (see Appendix~\ref{app:maximalset} for some more discussion about this point).

\section{Resource analysis of classes of pure states}\label{s:ps}
Given its relevance in the general framework just introduced, we now use the WLN to assess the resourcefulness of some paradigmatic examples of non-Gaussian states.
In particular, besides the aforementioned class of cubic phase states, we focus also on states that are of relevance in quantum optical experiments: photon-added, photon-subtracted, and cat states.

In addition to the WLN, given that we only consider pure states, we also calculate the non-Gaussianity [see Eq.~(\ref{eq:nong})].
As said, the latter is still not proved to be a monotone in our framework, however the comparison between the two quantities is particularly fruitful to single out the properties of the states considered.
\subsection{Cubic phase state}
\label{ssec:cps}
As recalled, a particularly important non-Gaussian continuous variable state is the so called cubic phase state~\cite{Gottesman2001}. For finite squeezing it is defined as
\begin{equation}
|\gamma , r \rangle = \exp \left[ i \gamma \hat{x}^3 \right] \hat{S}(r) |0 \rangle,
\end{equation}
where the squeezing operator $\hat{S}(r)=\exp\left[-\frac{i}{2} r \left( \hat{x} \hat{p} + \hat{p} \hat{x} \right) \right]$ for $r>1$ squeezes in momentum and anti-squeezes in position --- \textit{i.e.}, the Heisenberg evolution of the position operator is $\hat{S}(r)^\dag \hat{x} \hat{S}(r) = e^r \hat{x}$.
This implies that a squeezing unitary can be used to change the value of $\gamma$ of a cubic phase \emph{gate}~\cite{Gottesman2001}:
\begin{equation}
\label{eq:squeezing_cubic_gate}
\hat{S}(r)^\dag \exp\left[ i \gamma \hat{x}^3 \right] \hat{S}(r) = \exp\left[ i \gamma e^{3 r} \hat{x}^3 \right].
\end{equation}

This identity shows that we can ``consume'' the initial squeezing to enhance the nonlinear parameter by anti-squeezing the state (a Gaussian unitary)
\begin{equation}
\label{eq:squeezing_cubic}
| e^{3 r'} \gamma , r \rangle = S(-r') | \gamma , r + r' \rangle.
\end{equation}
This means that every monotone must be a function of the effective parameter $e^{3r} \gamma$, since it has to be invariant under Gaussian unitaries:
\begin{equation}
\mathcal{M} \left( | \gamma , r \rangle \right) = \mathcal{M} \left( | e^{3 r} \gamma , 0 \rangle \right) = f\left( e^{3 r} \gamma  \right)\;.
\label{eq:cubic_mono}
\end{equation}
As a consequence the contour lines of any monotone on the plane $(r,\gamma)$ are of the form $\gamma \propto e^{-3 r}$. In particular, Eq.~(\ref{eq:cubic_mono}) shows that the resourcefulness of the cubic phase state can be boosted by increasing the initial squeezing.

We remark that in the case of infinite squeezing $r\to \infty$ Eq.~\eqref{eq:squeezing_cubic} formally means that we can freely interconvert between ideal cubic phase states with simple Gaussian operations. This is consistent as long we assume to be in the degenerate case where the monotone assumes an infinite value for every cubic phase state, irregardless of the value of $\gamma$.

For a pure cubic phase state we can also compute the relative entropy of non-Gaussianity~\eqref{eq:nong}, which is again invariant for Gaussian unitaries
\begin{equation}
\delta \left[ | \gamma, r \rangle \right] = h\left(\sqrt{1+9 \left( e^{3 r} \gamma \right)^2 }\right),
\end{equation}
where $h(x) = \left( \frac{x+1}{2} \right) \log \left( \frac{x+1}{2} \right) - \left( \frac{x-1}{2} \right) \log \left( \frac{x-1}{2} \right) $; we can explicitly see the dependence on the combination $e^{3 r} \gamma$.
This measure goes to infinity as $\log \left(  e^{3 r } \gamma \right)$ for $ e^{3r} \gamma \to \infty$, as expected.

\begin{figure}[h!]
\includegraphics[width=\columnwidth]{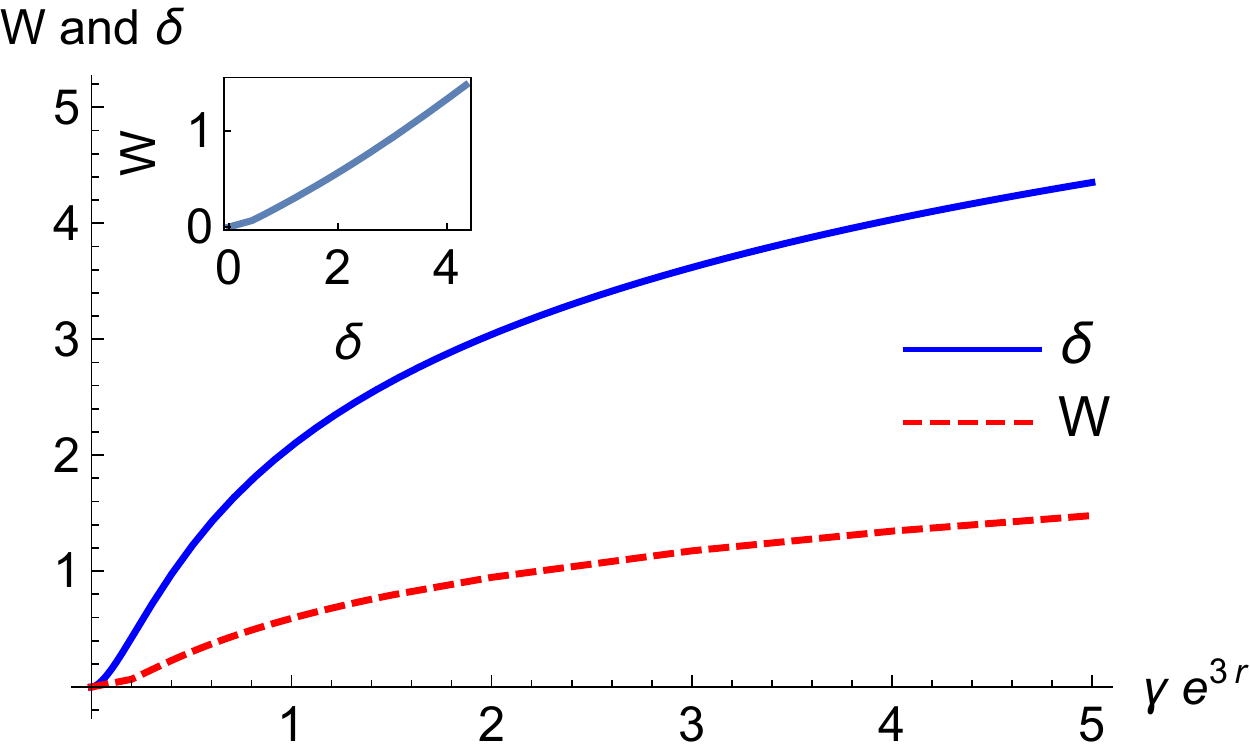}
\caption{(Color online) Non-Gaussianity $\delta$ (solid blue) and WLN $\mathsf{W}$ (dashed red) of the cubic phase state as a function of their unique parameter $\gamma e^{3r}$.
Inset: parametric plot of the two quantities.}\label{fig:cubic_mana_ng}
\end{figure}

We are working with pure states and therefore the Hudson theorem implies that if one measure is zero also the other has to be zero.
Furthermore, in this and in the following examples we observe that, as long as both the WLN $\mathsf{W}$ and the non-Gaussianity $\delta$ are functions of a single effective parameter, the two measures are monotonic and thus display the same qualitative behavior.
We remark that the same fact has also been observed for ground states of anharmonic potentials~\cite{Albarelli2016a}.
Given this heuristic argument, we also expect the WLN of the cubic phase state to be a monotonically increasing function of its effective parameter, with a behavior similar to the measure $\delta$; this is indeed what we observe from a numerical evaluation~\footnote{See Ref.~\cite{Brunelli2018} for an analytic expression of the Wigner function.}, see Fig.~\ref{fig:cubic_mana_ng}.
In particular we expect it to diverge like the non-Gaussianity monotone in the limit of infinite squeezing or nonlinearity, in accordance to the intuition from Eq.~\eqref{eq:squeezing_cubic}.

\subsection{Photon subtracted/added Gaussian states}
The single-mode photon subtracted and photon added Gaussian states are respectively defined as $|\alpha,r \rangle_\mathsf{sub} =N_\mathsf{sub}^{-1/2} \, \hat{a} D(\alpha) S(r) |0 \rangle$ and $|\alpha,r \rangle_\mathsf{add} =  N_\mathsf{add}^{-1/2} \, \hat{a}^\dag D(\alpha) S(r) |0 \rangle$, where $N_\mathsf{sub} = \sinh^2 r + |\alpha|^2$ and $N_\mathsf{add} = 1 + \sinh^2 r + |\alpha|^2 $ are normalization constants. These states have been realized experimentally \cite{Ourjoumtsev2006,NeergaardNielsen2006,Ourjoumtsev2007a,Allevi2010,Allevi2010b,Barbieri2010} and they have recently been suggested as non-Gaussian ancillas to implement arbitrary non-Gaussian operations~\cite{Arzani2017}. Multimode photon subtracted and added Gaussian states have also shown a nontrivial interplay with entanglement~\cite{Walschaers2017}.

We can employ again the invariance under Gaussian operations to get 
\begin{align}
\mathcal{M}\left[ |\alpha,r \rangle_\mathsf{sub} \right] &= \mathcal{M} \left[ N_\mathsf{sub}^{-1/2} \left( e^{i \psi} \sinh |r| |1 \rangle + \alpha |0\rangle \right) \right] \label{eq:sub_mono} \\
\mathcal{M}\left[ |\alpha,r \rangle_\mathsf{add} \right] &= \mathcal{M} \left[ N_\mathsf{add}^{-1/2} \left( \cosh | r|  |1 \rangle + \alpha^* |0 \rangle \right) \right],\label{eq:add_mono}
\end{align}
where $r=|r| e^{i \psi}$ and $\mathcal{M}$ represents a generic monotone.

\begin{figure}[h!]
\includegraphics[width=\columnwidth]{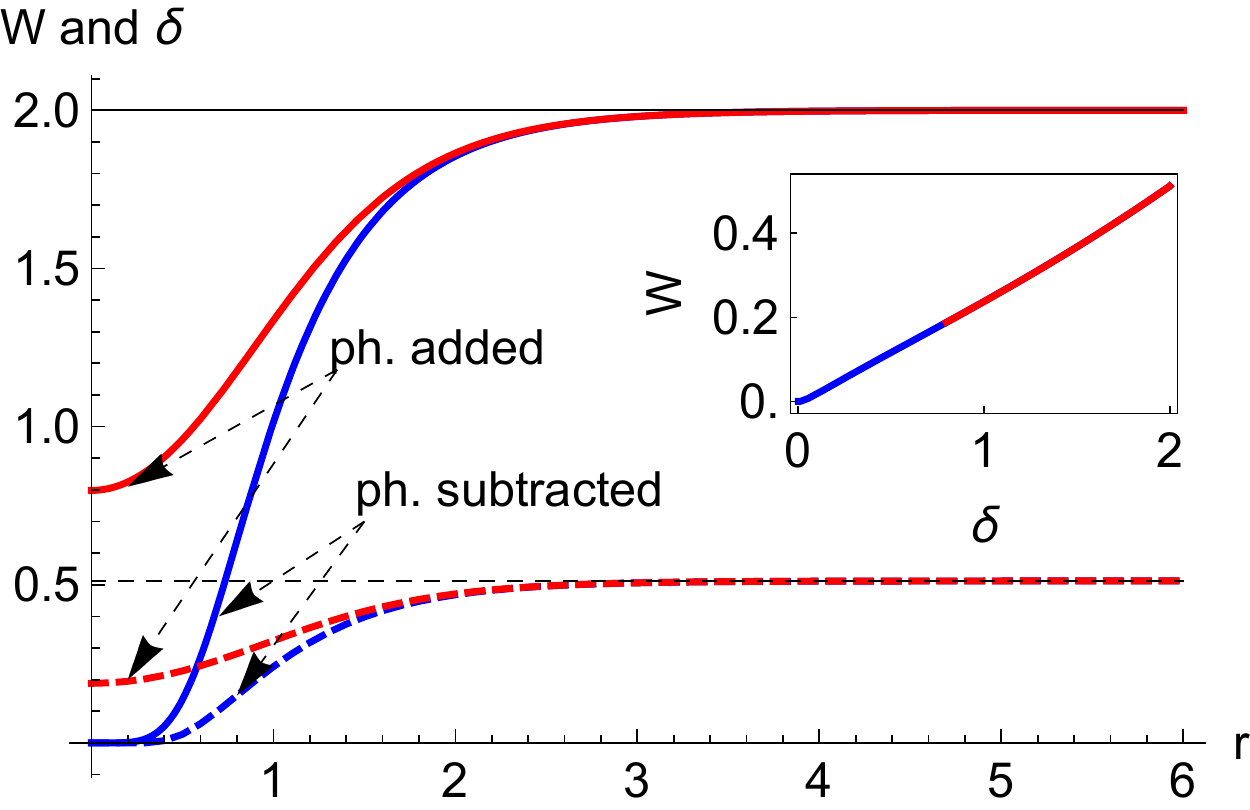}
\caption{(Color online) Non-Gaussianity $\delta$ (full lines) and WLN $\mathsf{W}$ (dashed lines) of photon subtracted (blue, lower curves) and photon added (red, upper curves) states as a function of $r$ for fixed $|\alpha | = 1$.
The horizontal black lines represent the value of the two figure of merit for the state $|1 \rangle$.
Inset: parametric plot of the two quantities. Since a photon-added state is non-Gaussian for any value of $r$, both $\mathsf{W}$ and $\delta$ never go to zero. In the region $\delta \gtrapprox 0.8$ the two parametric curves perfectly overlap.}\label{fig:add_sub_mana_ng}
\end{figure}

The results above suggest that the maximum amount of resource reachable by these two classes of states is that of a single photon state $|1\rangle$, a result in agreement with the physical intuition about the preparation of these states. Photon subtracted states can be prepared by sending the input state in a high-trasmissivity beam-splitter and then conditioning on a single photon detection on an output mode. On the other hand, photon addition can be implemented as beam splitting the input state with a single photon state, and then conditioning the output on the detection of no photons. This resource theoretical analysis shows that measurements and ancillary states are indeed equivalent resources in this case, as clearly confirmed by the plot in Fig.~\ref{fig:add_sub_mana_ng}.
We remark that while these schemes are appropriate for single mode states, more complicated schemes might be needed for multimode states, see e.g.~\cite{Averchenko2014,Ra2017} for photon subtraction.

We can compute the non-Gaussianity~\eqref{eq:nong} for these pure states
\begin{align}
\delta \left[ |\alpha,r \rangle_\mathsf{sub} \right] = h \left( \sqrt{\frac{8}{\left(\left| \alpha \right| ^2 \text{csch}^2(r)+1\right)^3}+1} \right) \\ 
\delta \left[ |\alpha,r \rangle_\mathsf{add} \right] = 
h\left(\sqrt{\frac{8}{\left(\left| \alpha \right| ^2 \text{sech}^2(r)+1\right)^3}+1}\right);
\end{align}
once again this is a function of a single parameter in both cases.
In Fig.~\ref{fig:add_sub_mana_ng} we also observe that non-Gaussianity and WLN have the same qualitative behaviour.

\subsection{Cat states}
We now want to complement the intuition we gained with the previous examples on a different class of non-Gaussian states: Schrödinger cat states.
We are going to see that the two figures of merit represented by Non-Gaussianity $\delta$ and WLN $\mathsf{W}$ can also display a qualitatively different behavior.

We define a cat state as the superposition of two coherent states $|\alpha \rangle$ and $|-\alpha \rangle$ and we keep both the amplitudes and the relative phase as parameters, as follows
\begin{equation}
| \psi (\alpha,\phi,\theta) \rangle = \frac{1}{\sqrt{K}} \left( \cos \phi |\alpha \rangle + \sin \phi e^{i \theta} | - \alpha \rangle \right),
\end{equation}
where $K$ is a normalization constant
\begin{equation}
K = 1 + \sin \left( 2 \phi \right) \cos \theta e^{-2 | \alpha |^2}.
\end{equation}

The non-Gaussianity $\delta\left[ | \psi (\alpha,\phi,\theta) \rangle \right]$ is not a function of the absolute value $|\alpha|$ only, but it depends on both angles; we do not report here the cumbersome analytical expression of this quantity.
A comparison between the two figures of merit shows that their behavior is qualitatively the same as a function of $\phi$ and $\theta$, while they show a remarkable difference as functions of $|\alpha|$.
As a matter of fact, while the WLN is known to saturate to a finite value for increasing separation between the two Gaussian peaks of the Wigner function~\cite{Kenfack2004}, the non-Gaussianity diverges, i.e. $\lim_{|\alpha| \to \infty} \delta\left[ | \psi (\alpha,\phi,\theta) \rangle \right] = \infty$.

\begin{figure}[h!]
\includegraphics[width=\columnwidth]{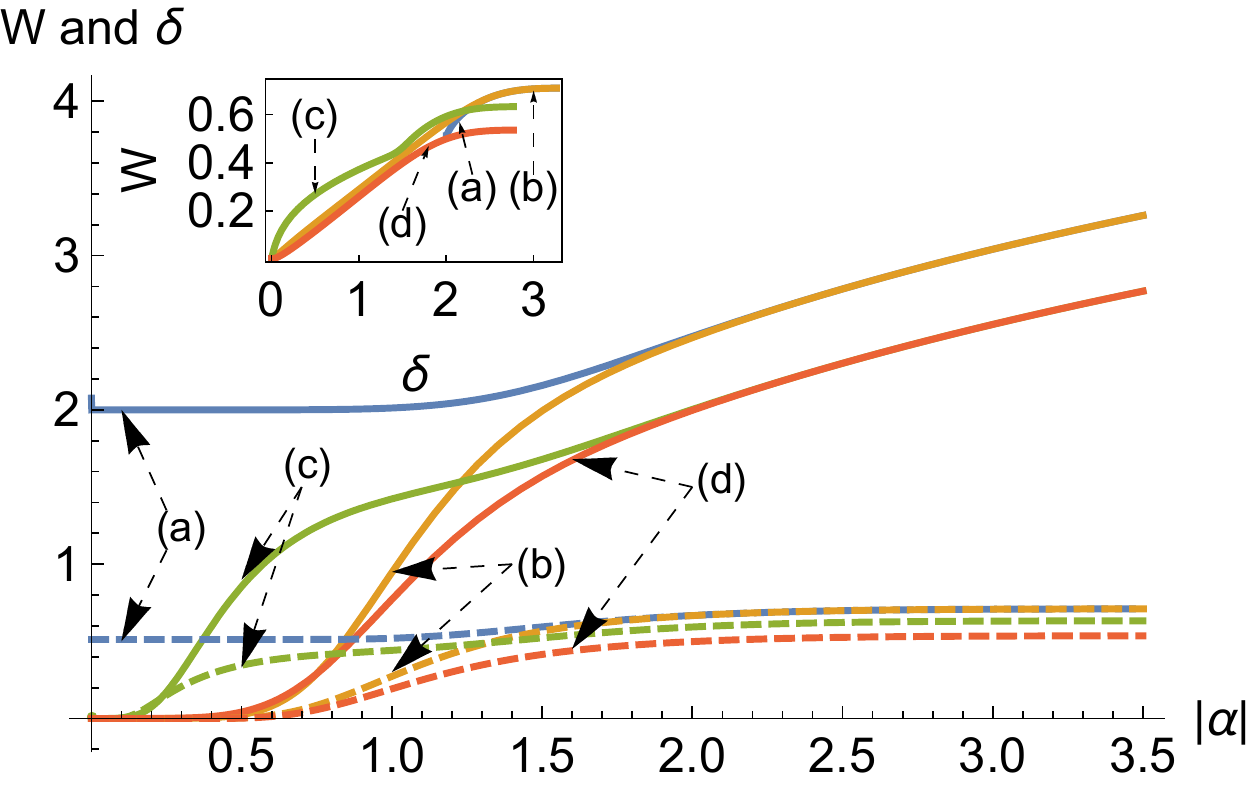}
\caption{
(Color online)
Non-Gaussianity $\delta$ (full lines) and WLN $\mathsf{W}$ (dashed lines) of cat states for different values of the parameters as a function of $|\alpha|$.
The values of the parameters are (a): $\phi = \pi /4, \theta = \pi$, (b): $\phi = \pi /4, \theta = 0$, (c): $\phi = \pi /8, \theta = \pi$ and (d): $\phi = \pi / 8, \theta = 0$.
Inset: parametric plot of the same quantities.}\label{fig:cat_mana_ng}
\end{figure}

This is shown in Fig.~\ref{fig:cat_mana_ng}, where we present the two quantities for a choice of parameters $\phi$ and $\theta$ as a function of $|\alpha|$.
We stress that even though the two quantities have a different behavior, they still remain monotonically increasing functions of one another (but not \emph{strictly} monotonic). It is reasonable to ascribe this difference to the fact that non-Gaussianity is sensitive to the distance between the state in question and pure Gaussian states. Given the double-peaked structure of cat states such a distance is bounded to increase indefinitely for large energies. On the other hand the WLN is clearly insensitive to this.

\subsection{Comparison at fixed energy}
We conclude this section by studying the behavior of the monotones as a function of the mean number of bosonic excitations $\langle \hat{n} \rangle$, a standard resource in CV quantum information processing.
In general there is no one to one relationship between mean energy and any QnG or WN monotone, therefore we consider the maximum value of the monotones for every fixed value of $\langle \hat{n} \rangle$.

For the cubic phase state the problem amounts to maximizing the parameter $e^{3r}\gamma$.
We find that the WLN of the cubic phase state is a monotonically increasing function of $\langle \hat{n} \rangle$ as intuitively expected.
For photon subtracted and photon added Gaussian states the problem amounts to maximizing the probability of the $|1\rangle$ component in~\eqref{eq:sub_mono} and~\eqref{eq:add_mono}.
When $\langle \hat{n} \rangle > 1$ the maximum value of both monotones is the same as for the state $|1 \rangle$ and this is achieved for $\alpha = 0$.
The photon-subtracted state can also have $\langle \hat{n} \rangle \leq 1$, in this case the maximal value for both monotones is equivalent to that of the state $\sqrt{1 - \langle \hat{n}  \rangle } |0\rangle + \sqrt{\langle \hat{n} \rangle} |1 \rangle$~\footnote{We remark that the maximum value of the monotones is achieved for a vanishing value of the parameters $\alpha$ and $r$ and this corresponds to the limit of a vanishing probability of a successful photon subtraction, i.e. the normalization constant $N_\mathsf{sub} \to 0$}.
For cat states we restrict to equal amplitudes of the two components, i.e. $\phi=\pi/4$.
In this case, for $\langle \hat{n} \rangle > 1$ we have that $\theta = \pi$, i.e. the odd cat state, is always optimal.
However, when $\langle \hat{n} \rangle < 1$ the state with $\theta = \pi$ does not exist and we need to numerically find the best angle $\theta$ for every value of $\langle \hat{n} \rangle$.

\begin{figure}[h!]
\includegraphics[width=\columnwidth]{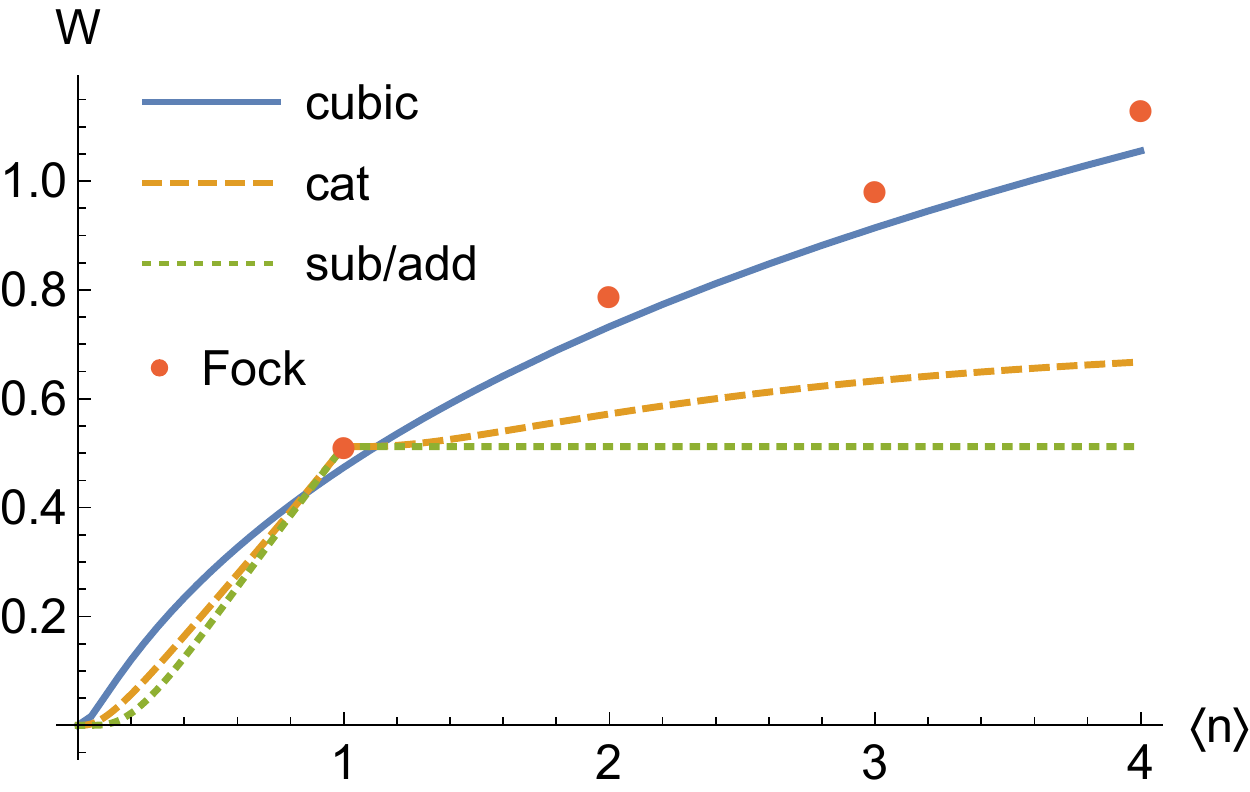}
\caption{
(Color online)
Maximal value of the Wigner logarithmic negativity as a function of the mean number of bosonic excitations for the considered classes of states, the solid light blue line represents the cubic phase state, the dashed orange line the class of cat states (with equal amplitude) and the dotted green line represents photon subtracted and added Gaussian states, the red dots represents Fock states.
}\label{fig:mana_energy}
\end{figure}

All these findings are summed up in Fig.~\ref{fig:mana_energy}, where we report an explicit comparison of the WLN $\mathsf{W}$ as a function of $\langle \hat{n} \rangle$.
We also show points corresponding to Fock states $|n\rangle$, which have an higher value of $\mathsf{W}$ than the classes of states we consider.
In particular, for $\langle \hat{n}\rangle = 1$ photon subtracted/added Gaussian states and odd cat states reduce to the single photon Fock state $|1\rangle$. 

We remark that the same qualitative analysis applies also to the non-Gaussianity $\delta$; however, in this case it can be proven that Fock states have the maximum value of $\delta$ at fixed energy~\cite{Genoni2010}.

\section{Negativity Concentration via Passive Gaussian operations}
\label{sec:concentration}
We have already recalled that, given an arbitrarily large supply of cubic phase states, it is possible to generate any state via ideal GPs. It is therefore possible to increase the amount of quantum non-Gaussianity of a state, and in particular to distill WLN. Given the relevance of WLN established by our framework, it is relevant to consider experimentally realistic settings that can concentrate the amount of WLN via operational GPs. Specifically, taking inspiration from existing CV quantum state engineering protocols based on linear optics, we now want to study the task of concentrating the negativity of the Wigner function from many copies of an input state to a single output state. We know that when a state can be transformed to another via GPs, then the conversion rate is limited by the bound in Eq.~(\ref{eq:add_mon_bound}), where we use the WLN as monotone. 

A more general framework which also makes use of single photon sources and general Gaussian measurements detection has been presented in~\cite{Park2014} to implement arbitrary nonlinear potentials.
Some general calculations about states obtainable starting from Fock states by applying two-mode interactions and conditional operations based on homodyne post-selection can be found in~\cite{Ghose2006}.

\subsection{Quantitative study of a negativity concentration protocol}
Now we exploit the theoretical framework described in the previous section to analyze a quantum state engineering protocol based on beam splitter interaction.
In what follows we consider the WLN $\mathsf{W}$ defined in Eq.~\eqref{eq:CV_mana} as the resource monotone.
Each probabilistic protocol is indeed a GP $\Lambda$ that converts $k$ copies of a resource state $\varrho$, into $m$ copies of a state $\sigma$, with a given probability $p$. 

Since we focus on the ``negativity concentration'' properties of these protocols, we introduce two figures of merit.
The first one corresponds to the resource gain of the protocol, and it is defined as the relative difference between the WLN of the output state and the WLN of the input state,
\begin{align}
\epsilon[\Lambda] = \frac{\mathsf{W}(\sigma) - \mathsf{W}(\varrho)}{\mathsf{W}(\varrho)}.
\end{align}
The second figure of merit quantifies the efficiency of the protocol, and it is defined as
\begin{align}
\eta[\Lambda] = p \, \frac{m \mathsf{W}(\sigma)}{k \mathsf{W}(\varrho)}  \leq 1 \,,
\end{align}
whose maximum value is one, as a consequence of Eq. (\ref{eq:eq:resbound_additive}).\\

\begin{figure}[h!]
\includegraphics[width=.7\columnwidth]{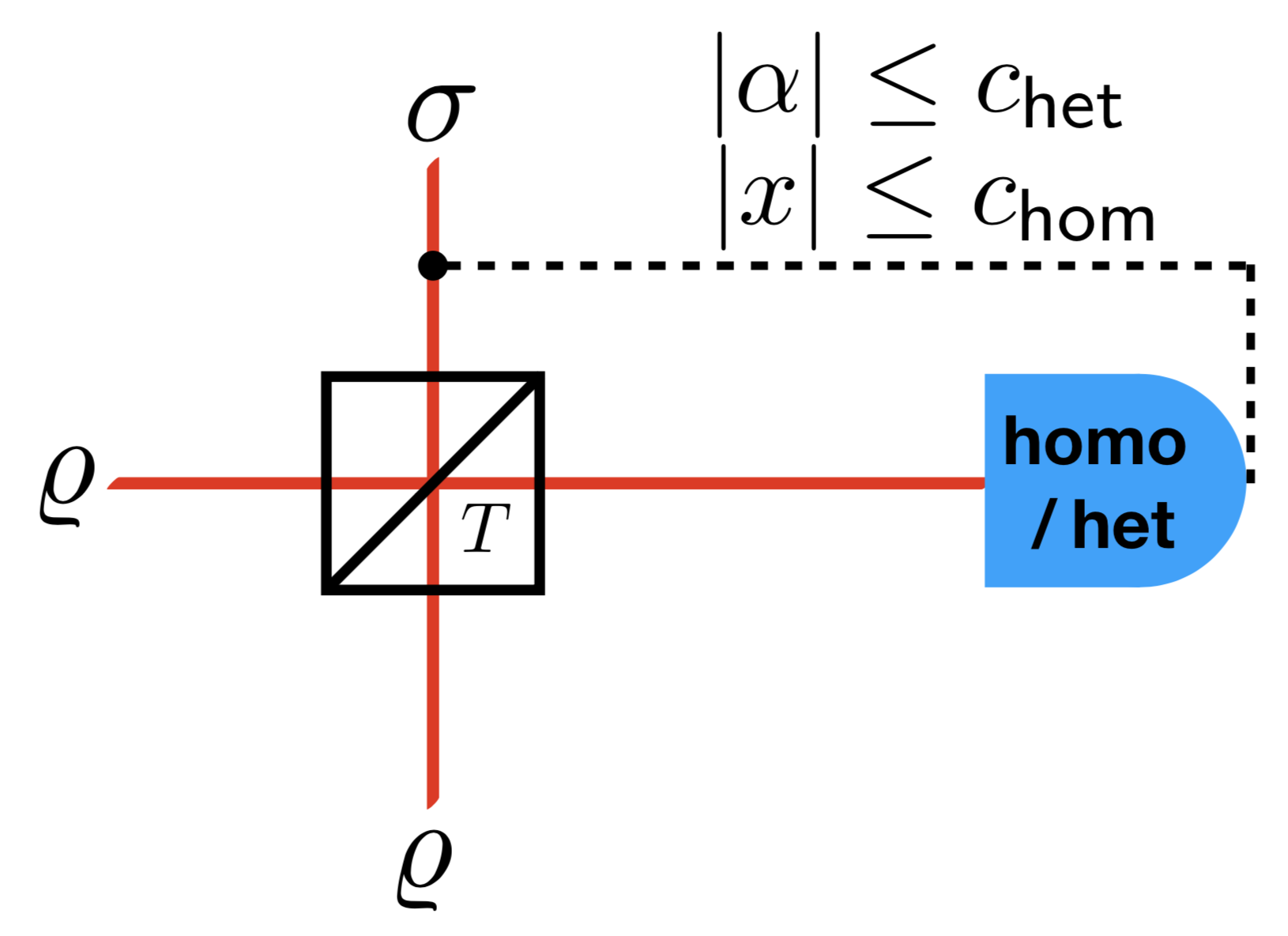}
\caption{Schematic representation of the concentration protocol.
The detector represents either an homodyne (``homo'') with outcome $x$ or an heterodyne one (``het'') with outcomes $\alpha = |\alpha| e^{i\theta}$.
The input states are two copies of $\varrho$, while the output state $\sigma$ is postselected according to the output values of the measurement.
}
\label{f:scheme}
\end{figure}

In particular, here we focus on a protocol that has been implemented experimentally as described in Refs. \cite{Etesse2014a,Etesse2015}: a pair of identical
Fock states $\varrho= |n\rangle\langle n|$ is mixed at a beam-splitter with transmissivity $T$, and a homodyne 
detection of the quadrature $\hat{x}$ is performed on one of the two arms, see Fig.~\ref{f:scheme}.
We also study the case of heterodyne detection for comparison.

To gain intuition about this protocol, consider the output state of the beam splitter, which is of the form $\sum_{p=0}^{2 n} f_p | p, 2n - p\rangle$. If we were able to condition on the detection of no photons we would obtain the state $|2n \rangle$, which in turn would imply the concentration of the resource. However, conditioning on the vacuum via Gaussian measurements occurs with zero probability and is therefore unfeasible.
The actual protocol works instead conditioning on outcomes that belong to a finite interval around zero. 
This produces a POVM element which is very close to the ideal projection on the vacuum $| 0 \rangle \langle 0 |$, but with non-zero probability and therefore feasible \cite{Etesse2014a,Etesse2015}.

In the case of heterodyne detection, the state on the other arm of the beam splitter is kept if the outcome of the measurement
is in the range $0 < |\alpha | < c_\mathsf{het}$, while we average over the phase $\theta$ (note that this makes the POVM elements diagonal in Fock basis).
Analogously, in the case of homodyne detection the output is conditioned on the result $x$ falling within  the interval $[-c_\mathsf{hom}, c_\mathsf{hom}]$.
This means that the output states are the following mixed states:
\begin{align}
\sigma_\mathsf{het} &= \int_{0}^{c_\mathsf{het}} |\alpha| \mathrm{d} |\alpha|  \int_{0}^{2 \pi} \frac{\mathrm{d} \theta}{\pi} \, \frac{ \hbox{Tr}_b[ U_\mathsf{bs} \left( \varrho \otimes \varrho \right) U_\mathsf{bs}^\dag \left( \id \otimes | \alpha \rangle\langle \alpha | \right)] }{p_\mathsf{het}} \,  , \\
\sigma_\mathsf{hom} &= \int_{-c_\mathsf{hom}}^{c_\mathsf{hom}} \mathrm{d} x \, \frac{ \hbox{Tr}_b[ U_\mathsf{bs} \left( \varrho \otimes \varrho \right) U_\mathsf{bs}^\dag \left( \id \otimes |x\rangle\langle x| \right) ] }{p_\mathsf{hom}} \, ,
\end{align}
where the probabilities of success of the protocol is 
\begin{align}
p_\mathsf{het} &=  \int_{0}^{c_\mathsf{het}} |\alpha| \mathrm{d} |\alpha|  \int_{0}^{2 \pi} \frac{\mathrm{d} \theta}{\pi} \, \,  \hbox{Tr}_{ab}[ U_\mathsf{bs} \left( \varrho \otimes \varrho \right) U_\mathsf{bs}^\dag \left( \id \otimes | \alpha \rangle\langle \alpha | \right)] \,, \\
p_\mathsf{hom} &=  \int_{-c_\mathsf{hom}}^{c_\mathsf{hom}}  \mathrm{d} x \,  \hbox{Tr}_{ab}[ U_\mathsf{bs} \left( \varrho \otimes \varrho \right) U_\mathsf{bs}^\dag \left( \id \otimes |x\rangle\langle x| \right)] \,.
\end{align}

With homodyne conditioning it is not possible in general to obtain the state $|2 n \rangle$, since we would need to condition around a value $x$ where all the wavefunction $\langle x | n \rangle=0$ for $n > 1 $.
This is possible just in the case of single photon states since the superposition is simply between $|0,2 \rangle$ and $|2,0\rangle$ and therefore we could approximate a projection on the vacuum by conditioning the homodyne detection in a region around the zero of $\langle x | 2 \rangle$, which is obtained for $x_0 = 1 / \sqrt{2}$.

\begin{figure}[h!]
\centering
\includegraphics[width=.95\columnwidth]{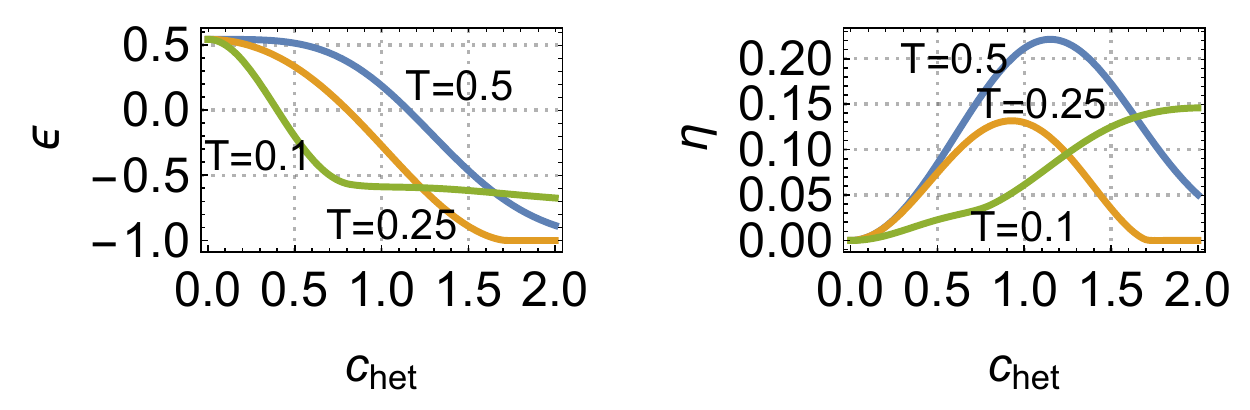}\\
\vspace{2mm}
\includegraphics[width=.95\columnwidth]{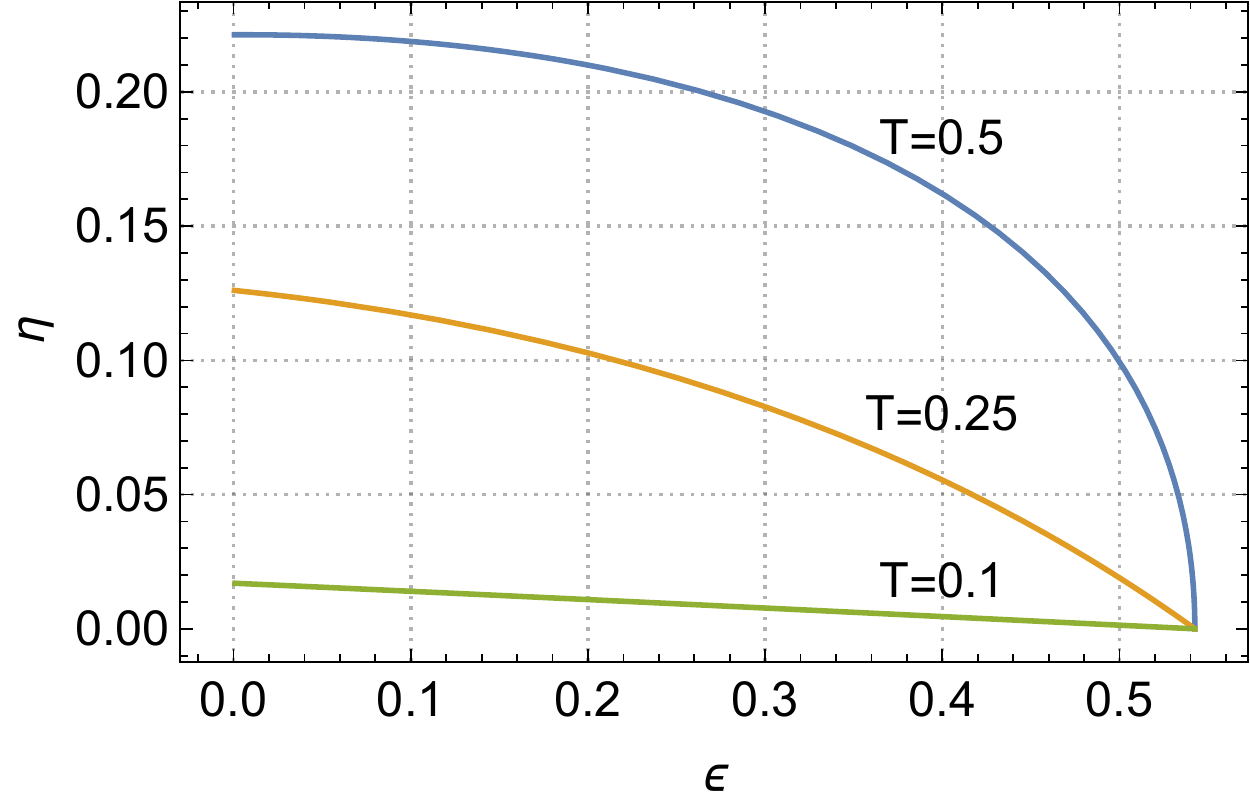}
\caption{
(Color online)
Efficiency $\eta$ and negativity gain $\epsilon$ for the heterodyne concentration scheme with two single photon Fock states $|1 \rangle$ in input as a function of the conditioning parameter $c_\mathsf{het}$ for different beam-splitter transmissivities.}
\label{f:Fock1PureAhetero}
\end{figure}

\begin{figure}[h!]
\includegraphics[width=.95\columnwidth]{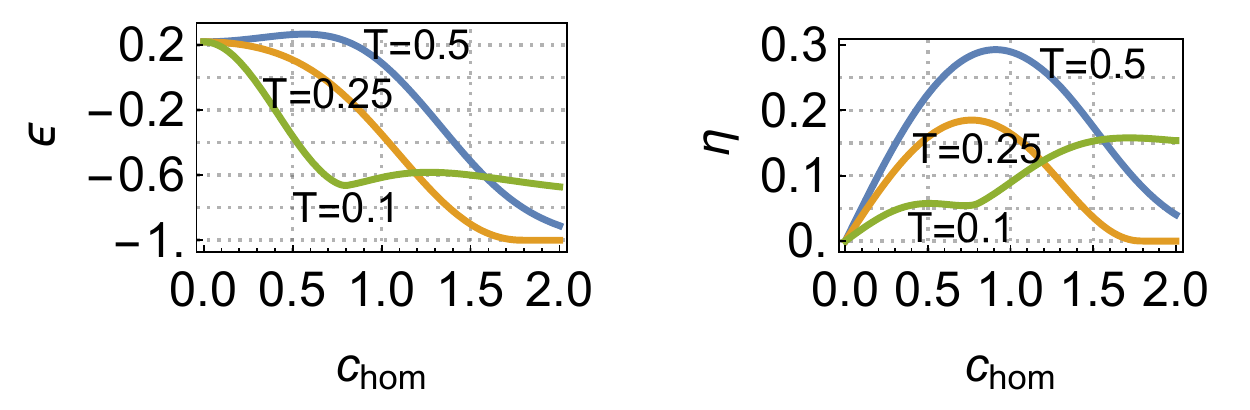}
\\
\vspace{2mm}
\includegraphics[width=.95\columnwidth]{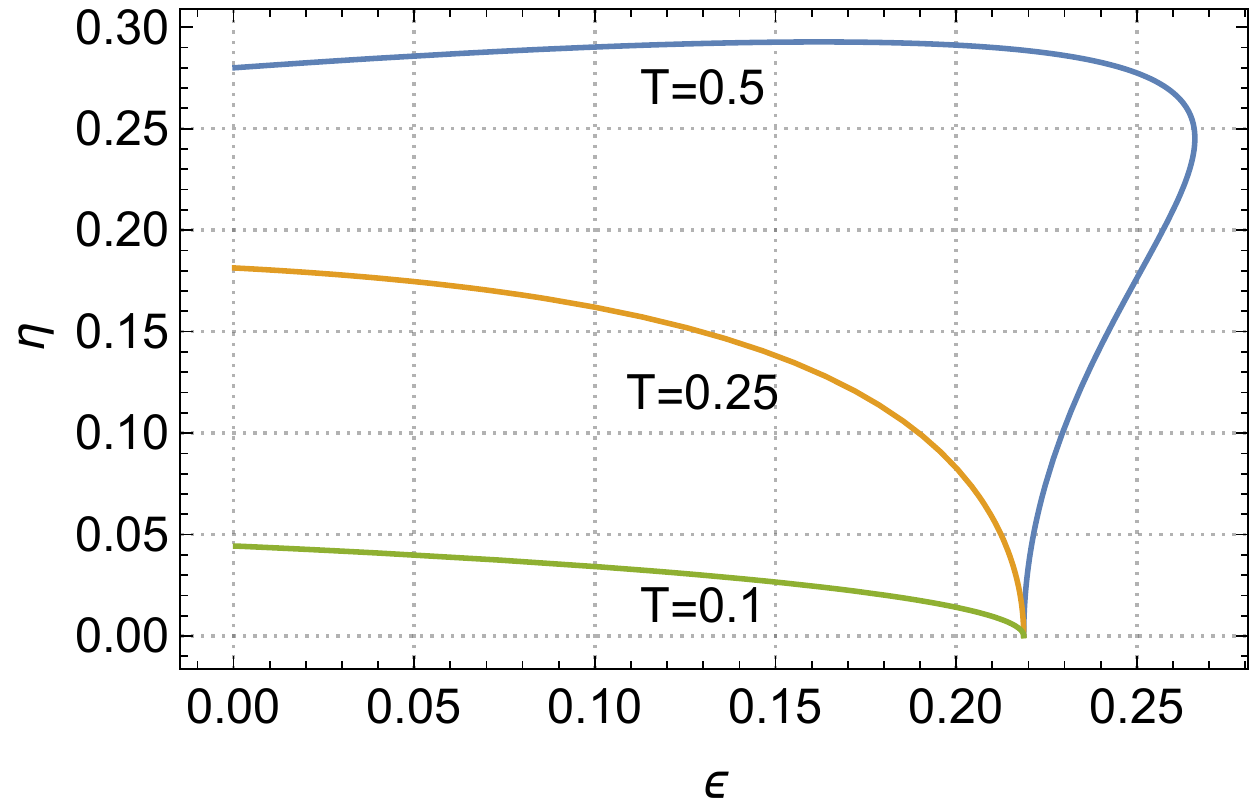}
\caption{(Color online) Efficiency $\eta$ and negativity gain $\epsilon$ for the homodyne concentration scheme with two single photon Fock states $|1 \rangle$ in input.
Top: both quantities as a function of the conditioning parameter  $c_\mathsf{hom}$ for different beam-splitter transmissivities.
Bottom: parametric plot of efficiency and negativity when $c_\mathsf{hom}$ is varied. 
}
\label{f:Fock1PureAhomo}
\end{figure}

In terms of our figures of merits we have $k=2$ input and $m=1$ output copies.
To start, we study the protocol with two single photon states $|1 \rangle$ in input.
The protocol efficiency $\eta[\Lambda]$ and the resource gain $\epsilon[\Lambda]$ are plotted in the upper panels of Fig. \ref{f:Fock1PureAhetero} and \ref{f:Fock1PureAhomo}, as a function of the heralding parameters $c_\mathsf{het}$ and $c_\mathsf{hom}$, and for different values of $T$.
We observe that $\epsilon \geq 0$ up to certain threshold values of $c_\mathsf{het}$ and $c_\mathsf{hom}$, depending on the the transmissivity $T$.
On the other hand the maximum efficiency can always be achieved for a balanced beam-splitter, since this maximizes the matrix element $\langle 2 | \sigma |2 \rangle$ of the output states.
Typically, these \emph{optimal} values are obtained for values of $c_\mathsf{het}$ and $c_\mathsf{hom}$ that in turn corresponds to lower values of $\epsilon$.
This tradeoff is better highlighted in the lower panels of Fig. \ref{f:Fock1PureAhetero} and \ref{f:Fock1PureAhomo}, where we show the efficiency $\eta$ as a function of the gain $\epsilon$.

We observe that we can indeed gain more negativity with heterodyne heralding, since we are approximately creating the state $|2 \rangle$, in this case there is a strict trade-off between $\eta$ and $\epsilon$, i.e. $\eta$ is a monotonically decreasing function of $\epsilon$.

For homodyne heralding we gain less negativity, but at least for $T=0.5$, there is an \emph{optimal} region, where \emph{large} values of the two figure of merits can be achieved (\emph{large values} in the sense that they are close to the maximum values that one can achieve via this protocol, by changing the interval width $c_\mathsf{hom}$).
However, by changing the transmissivity of the beam splitter we observe a monotonically decreasing behaviour of $\eta$ as a function of $\epsilon$, similarly to the heterodyne case.\\

\begin{figure}[h!]
\includegraphics[width=.95\columnwidth]{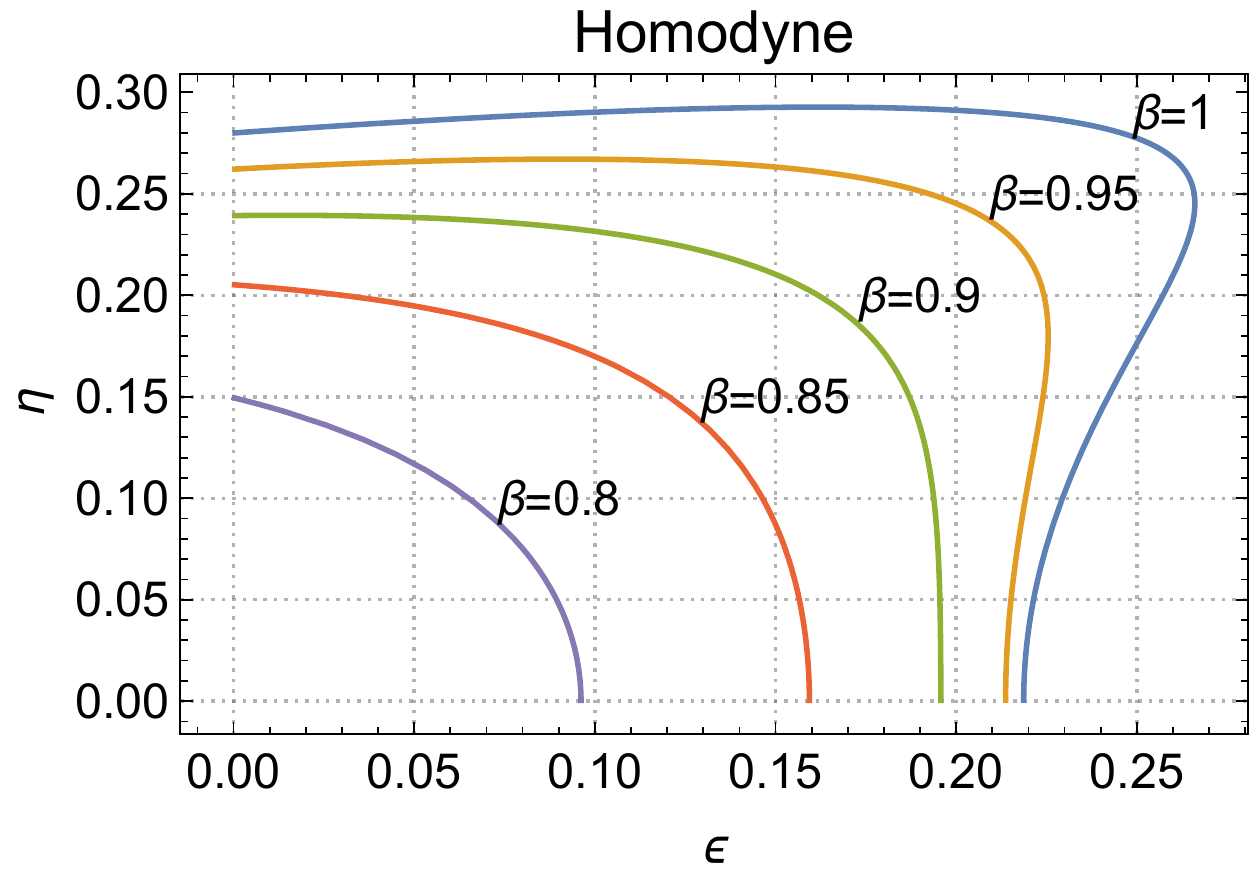}\\
\vspace{2mm}
\includegraphics[width=.95\columnwidth]{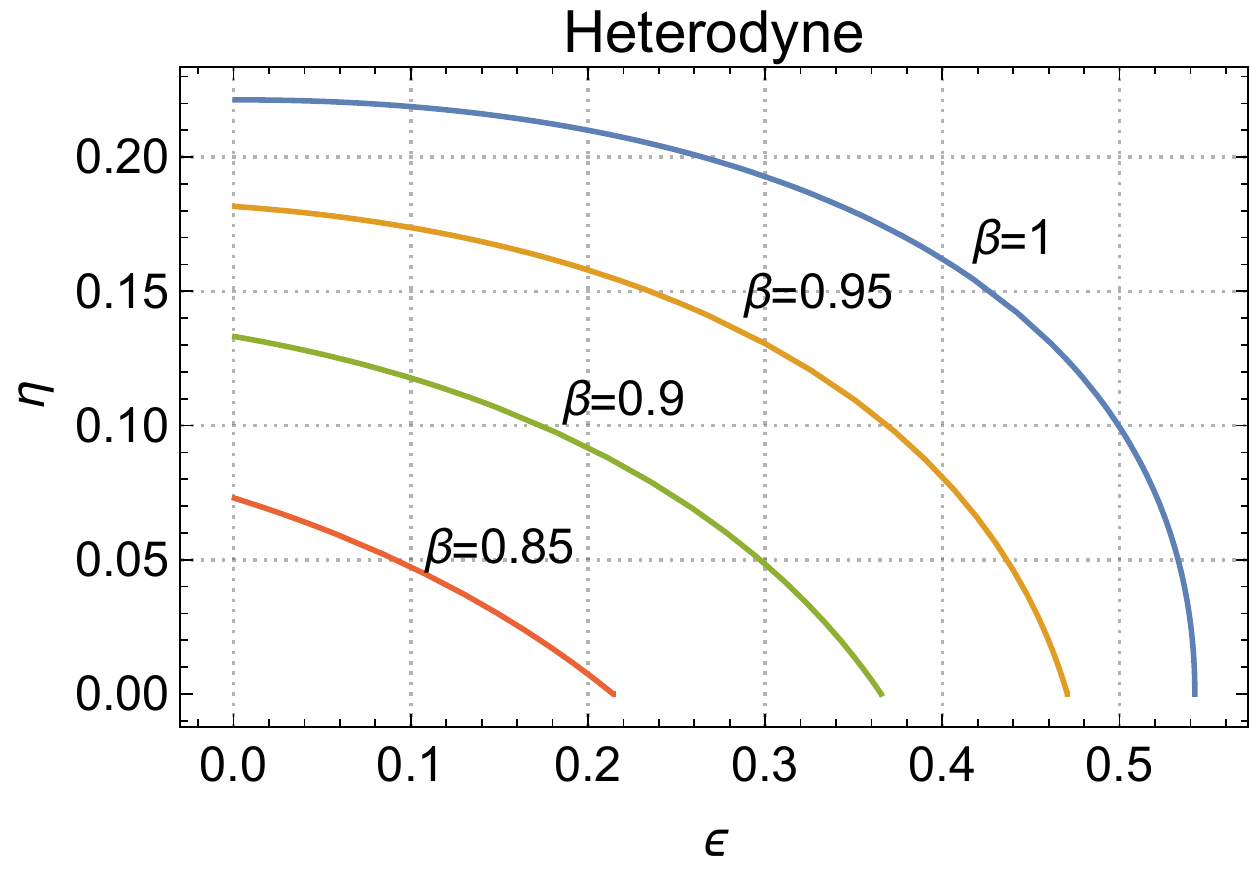}
\caption{(Color online) Parametetric plot of the efficiency $\eta$ and the negativity gain $\epsilon$ for both the heterodyne and homodyne protocols with two lossy single photon states $(1-\beta) | 0 \rangle \langle 0 | + \beta |1 \rangle \langle 1| $ for different values of $\beta$.}\label{f:Fock1MixedParametric}
\end{figure}

We then consider the same protocol with {\em lossy} Fock states $\varrho = \beta |1\rangle\langle 1| + (1-\beta) |0\rangle\langle 0|$ as input (resource) states and by fixing the transmissivity to $T=1/2$. We observe the same tradeoff in the parametric plot in Fig. \ref{f:Fock1MixedParametric}:  the protocol becomes less effective for decreasing values of $\beta$, and the {\em optimal} region for homodyne measurements is not present anymore for $\beta\lesssim 0.9$.\\

\begin{figure}[h!]
\includegraphics[width=.95\columnwidth]{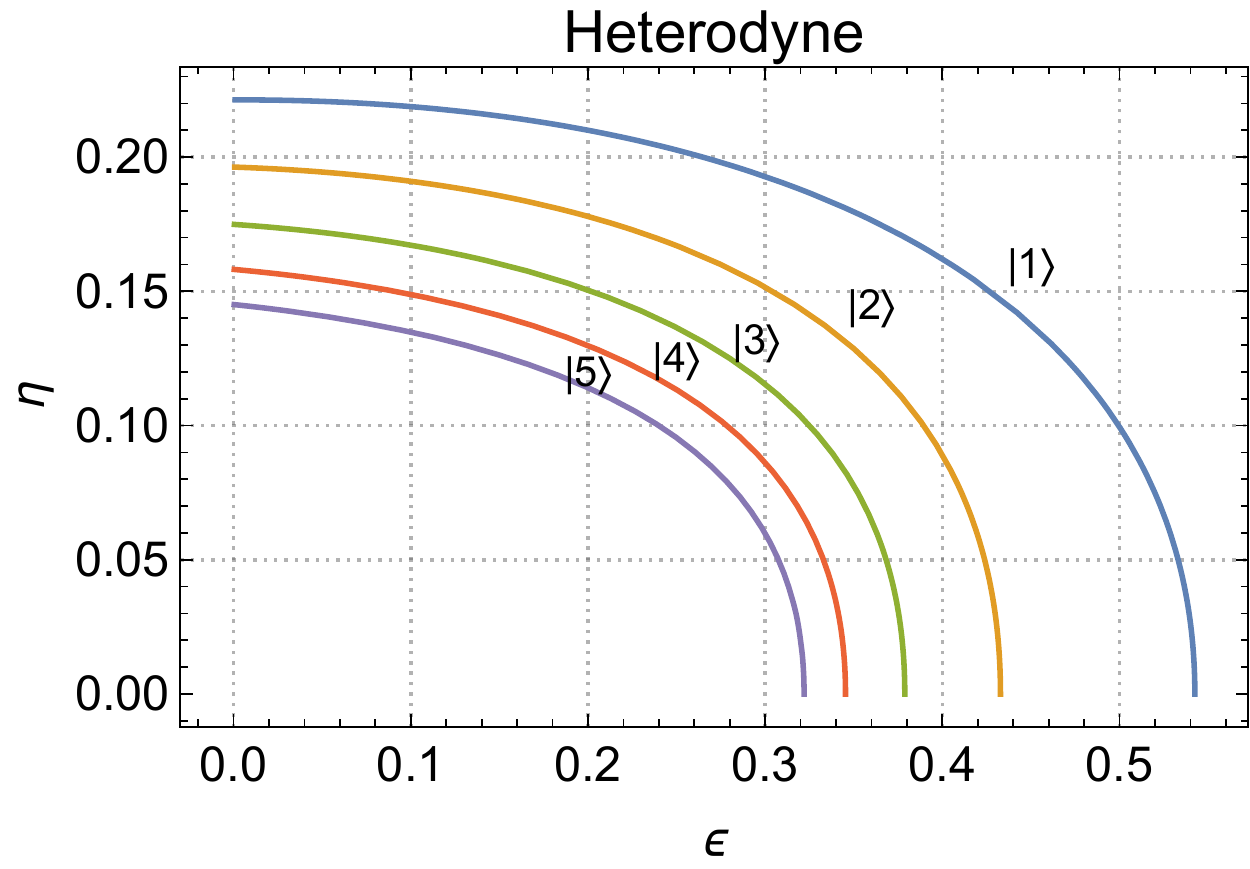}
\includegraphics[width=.95\columnwidth]{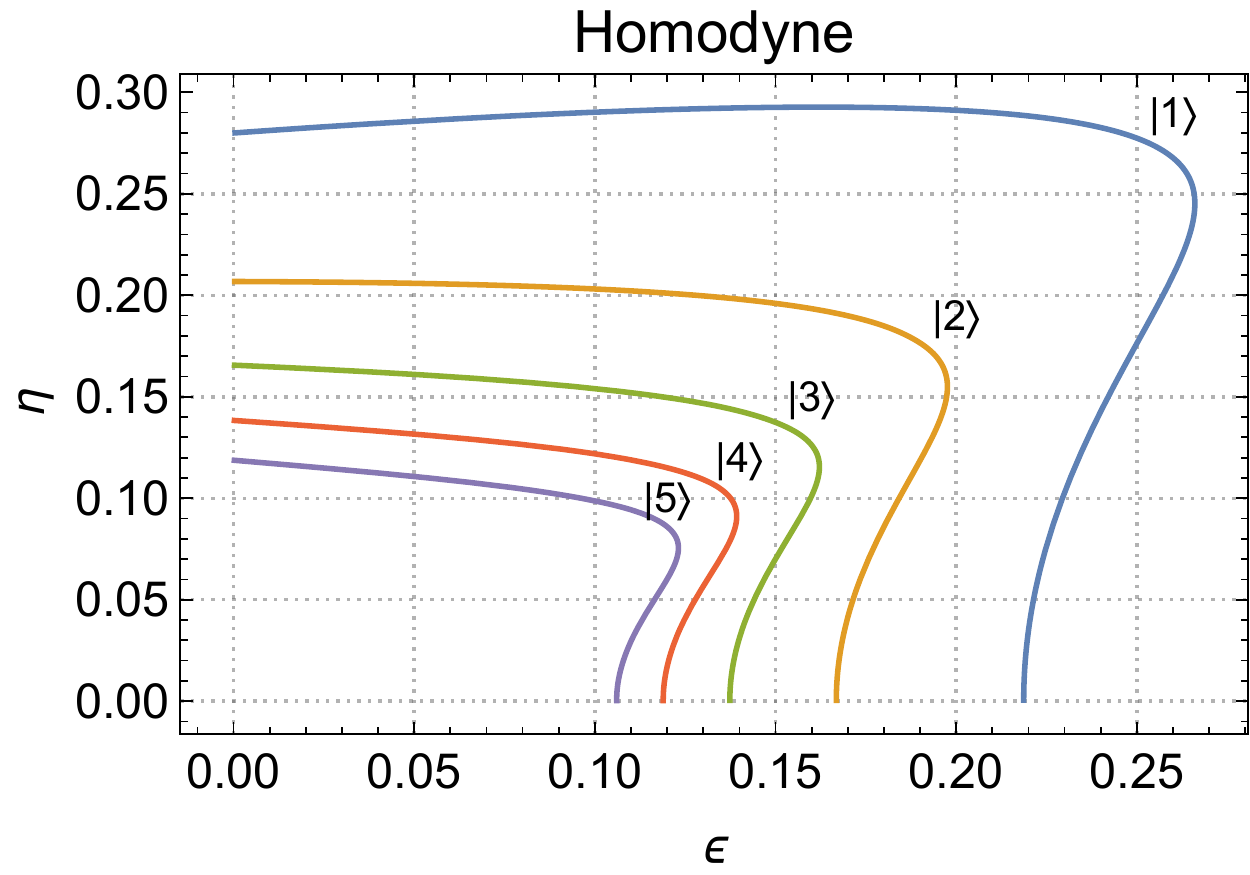}
\caption{(Color online) Parametetric plot of the efficiency $\eta$ and the negativity gain $\epsilon$ for both the heterodyne and homodyne protocols with two Fock states $|n \rangle $ in input, for $n=1,\dots,5$.
}\label{fig:FockNparametric}
\end{figure}

Finally, we consider Fock states $|n\rangle$ with $n$ up to $5$ as input resources (mixed at a balanced beam-splitter) and we plot in Fig.~\ref{fig:FockNparametric} $\eta$ as function of $\epsilon$. 
All the curves present a similar behaviour, but it is clear that the performances of the protocol in terms of our figures of merit decreases for increasing $n$. \\

\subsection{A remark about negativity concentration}
These schemes we studied are \emph{concentration} protocols at the single copy level, in the sense that the single outcome state is more resourceful than a single input state, but two input states must be used.

In particular, these schemes satisfy the following inequality
\begin{equation}
\frac{\mathsf{W} ( \sigma) }{\mathsf{W}(\rho^{\otimes 2})} < 1,
\end{equation}
which is a much stronger constraint than what imposed by the bound~\eqref{eq:eq:resbound1}, i.e. the WLN of the output state never surpass the WLN of the \emph{global} input state, not even probabilistically.

The reason for this bound is that negativity of Fock states is sublinear in $n$~\cite{Kenfack2004}, this implies that
\begin{equation}
\mathsf{W} \left( | 2 n \rangle \right) < 2 \mathsf{W} \left( | n \rangle \right)
\end{equation}
and the WLN of the output states is never higher than the one of the state $| 2 n \rangle$.

One might think that this is true in general for GPs, but this is not the case.
We found a simple counter-example to this by considering a two-mode non-Gaussian entangled input state of the form 
\[ | \psi_\mathsf{in} \rangle = \sqrt{1-\beta} |0 ,0 \rangle +  \sqrt{\beta} |1, 1 \rangle. \]
If one mode of this state is measured with an heterodyne detector and the outcome conditioned to be in certain intervals, the remaining state on the other mode is thus a mixture
\begin{align}
\tilde \sigma_\mathsf{out} = &\int_{\theta_1}^{\theta_2} \frac{\mathrm{d} \theta}{\pi} \int_{\alpha_1}^{\alpha_2} |\alpha| \mathrm{d} |\alpha| \nonumber \\
 &\biggl[ (1- \beta) |\langle \alpha | 0 \rangle|^2 | 0 \rangle \langle 0 | +  \beta |\langle \alpha | 1 \rangle |^2 |1 \rangle \langle 1 | + \nonumber \\
& + \sqrt{\beta (1-\beta)} \left( \langle \alpha | 0 \rangle \overline{\langle \alpha | 1 \rangle} |0 \rangle \langle 1| + \langle \alpha | 1 \rangle \overline{\langle \alpha | 0 \rangle} |1 \rangle \langle 0 | \right)
 \biggr]. \nonumber
\end{align}
For sufficiently small values of $\beta$ and conditioning ``far'' from the origin one can indeed obtain $\mathsf{W} ( \sigma_\textsf{out} ) > \mathsf{W} (| \psi_\mathsf{in} \rangle )$.
These outcomes are however very unlikely and the values for the efficiency $\eta$ are very low; even if the WLN is increased we are not close to the saturation of bound~\eqref{eq:eq:resbound1}.

For example we find that for $\alpha_1 = 2.5$, $\alpha_2 \to \infty$, $\theta_1 = - \pi / 6 $ and $\theta_2 = + \pi / 6 $ we have $\mathsf{W} ( \sigma_\textsf{out} ) -  \mathsf{W} (| \psi_\mathsf{in} \rangle ) \approx 0.05$, however the probability is very small $p_{\textsf{out}} \approx 5 \cdot 10^{-4} $ and therefore the efficiency is negligible as well $\eta \approx 8 \cdot 10^{-4}$.

\section{Conclusions}\label{s:out}
We have introduced a general and physically motivated framework for the resource theory of quantum non-Gaussianity and Wigner Negativity. 
After showing that no maximally resourceful state exists in this context, we defined a computable monotone ---the Wigner logarithmic negativity. We have used the latter to gauge both the resource content of experimentally relevant states and the efficiency of some resource concentration protocols.
\par
Our approach provides a reliable and robust set of tools to develop a resource theory based on the natural requirement of convexity for free states and operations.
As a matter of fact, notice that a resource theory of non-Gaussian operations has recently been introduced~\cite{Zhuang2018}, however dealing with the non-convex notion of non-Gaussianity. With the measure introduced there, a genuine non-Gaussian unitary (\textit{e.g.}, the self-Kerr interaction) can have the same (diverging) degree of non-Gaussianity as a simple statistical mixture of Gaussian channels. This fact highlights the importance of the operationally motivated framework that we introduced here. In addition, our framework also incorporates the peculiar feature of CV measurements that finite-probability output states are necessarily mixed. This is of practical relevance when assessing the interconversion rate of state manipulation protocols, as shown in Sec.~\ref{sec:concentration}. 
\par
We believe that understanding the possibilities and limits of manipulating non-Gaussian states with Gaussian operations is a challenging but timely topic of research. These manipulations --- that are becoming a reality in quantum optical experiments~\cite{Miwa2014} --- have been proved, for example, to counteract the effect of  decoherence of non-Gaussian states~\cite{Filip2013,Jeannic2017}.
In a similar fashion, protocols that require a non-Gaussian element can be often improved with Gaussian operations. This is the case of entanglement distillation based on photon subtraction, whose performances can be enhanced by implementing local Gaussian operations~\cite{Zhang2011,Fiurasek2011,Cernotik2012}.
\par
The results presented here promote a celebrated indicator of non-classicality --- the non-positivity of the Wigner function --- to a fully fledged monotone of the resource theory relevant for quantum information processing with infinite-dimensional systems. Indeed, we argue that the framework here introduced will contribute towards the development of sub-universal and universal CV quantum information processing.
\par
First, as recalled, the negativity of the Wigner function has been proven to be necessary for a quantum process to be not efficiently simulatable with classical phase-space techniques~\cite{Mari2012,Veitch2013,Pashayan2015,Rahimi-Keshari2015}.
In this context, the WLN here introduced represents a proper quantifier that was still missing. In addition, some sub-universal CV circuits have been rigorously proven hard to simulate classically. In particular, recent proposals based on non-Gaussian inputs and Gaussian operations and measurements~\cite{Lund2017a,Chakhmakhchyan2017a, Douce2016} fall clearly in the framework presented here --- which then provides quantitative tools for their analysis.
\par
Finally, regarding more general processing, non-Gaussian operations are known to be necessary to attain universality of state generation or computation.
Non-Gaussian operations can in turn be enabled by non-Gaussian states via gate teleportation, the Gaussian protocol at the basis of CV measurement-based quantum computation.
However, it is not clear in general which non-Gaussian states can play this crucial enabling role~\cite{Ghose2006,Etesse2014a,Vasconcelos2010,Arzani2017,Weigand2018,Sabapathy2018,Su2018} and at which cost in terms of circuit synthesis~\cite{Sefi2011}.
As progresses in analogous problems for DV systems have been made possible by the development of the resource theory for magic states \cite{Howard2014,Veitch2012a,Howard2017}, we expect similar advances to be possible in the context of CV systems with the assistance of the framework introduced here.

\emph{Note added} --- 
After the completion of this work, we have become aware of a related work by R.Takagi and Q. Zhuang~\cite{Takagi2018}.
\section*{Acknowledegements}
We acknowledge fruitful discussions with M. Brunelli, G.~Ferrini, O.~Houhou and M. Lostaglio.
MGAP is member of GNFM-INdAM.
MGG acknowledges support from Marie Skodowska-Curie Action H2020-MSCA-IF-2015 (project ConAQuMe, grant no. 701154).
AF acknowledges funding from the EPSRC (project EP/P00282X/1).

\appendix
\section{Phase-space formalism and Gaussian states}\label{app:phase_space}
Given $N$ bosonic modes we collect the $2N$ canonical self-adjoint operators as a vector $\hat{\vec{r}}=\left(\hat{q}_1, \hat{p}_1 , \dots ,  \hat{x}_N , \hat{p}_N \right)$, while the corresponding $N$ destruction operators are defined as $\hat{a}_j = \left( \hat{q}_j + i \hat{p}_j \right)/\sqrt{2}$; we also introduce the corresponding vector of classical phase space variables $\vec{r} = \left(q_1, p_1 , \dots ,  q_N , p_N \right)$.
The canonical bosonic commutation relations are $\left[ \hat{x}_i, \hat{p}_j\right] = i \delta_{ij}$ (we assume units such that $\hbar = 1$), or more compactly $\left[ \hat{\vec{r}},\hat{\vec{r}}^\mathsf{T} \right] = i \Omega$~\cite{serafini2017quantum}, where the canonical symplectic form $\Omega$ is defined as 
\begin{equation}
\Omega = \bigoplus_{i=1}^N \Omega_1, \qquad \Omega_1 = \begin{pmatrix}
0 &  1 \\
-1 & 0
\end{pmatrix}.
\end{equation}
The mapping between functions in phase space and Hilbert space operators is known as Weyl-Wigner transform~\cite{Case2008,Curtright2013}.
The Wigner transform of a generic operator on the infinite dimensional Hilbert space $N$ bosonic modes is defined as 
\begin{equation}
\mathcal{W}[\hat{O}](\vec{r})= \frac{1}{ \left( 2 \pi \right) ^{2 N} } \int_{\mathbb{R}^{2N}} \mathrm{d} \vec{v} \, e^{i \vec{v}^\mathsf{T} \Omega \vec{r}} \Tr \left[ \hat{O} e^{- i \vec{v}^\mathsf{T} \Omega \hat{\vec{r}}} \right],
\end{equation}
which explicitly shows the independence from the basis.
The Wigner transform contains all the information about the operator and in this sense it is a different representation of the same object.
The operators $\hat{D}_{\vec{r}} = e^{i \vec{r}^\mathsf{T} \Omega \hat{\vec{r}}}$ are the so called Weyl (or displacement) operators, while $\chi_{\hat{O}}(\vec{r}) = \Tr \left[ \hat{O} \hat{D}_{-\vec{r}} \right]$ is the the characteristic function.
If we evaluate the trace in position basis $\vec{\hat{q}} |\vec{s} \rangle_{q} = \vec{s} |\vec{s} \rangle_{q} $ we get the equivalent expression
\begin{equation}
\mathcal{W}[\hat{O}](\vec{q},\vec{p})= \frac{1}{\pi^N} \int_{\mathbb{R}^{N}} \mathrm{d}\vec{y} \, _q\langle \vec{q} + \vec{y} | \hat{O} | \vec{q}-\vec{y} \rangle_q \, e^{-2 i \vec{p} \cdot \vec{y}}.
\end{equation}
Without delving into mathematical details we simply assume that operators can be expressed as $\hat{O}=f(\hat{\vec{r}})$, where $f$ is a well-behaved enough function.

A fundamental property of the Wigner transform is that the trace of two operators can be expressed as an integral in phase space
\begin{equation}
\label{eq:tr_rule_ph_space}
\Tr[\hat{O}_1 \hat{O}_2] = (2 \pi)^N \int_{\mathbb{R}^{2N}} \mathrm{d} \vec{r} \mathcal{W}[ \hat{O}_1] ( \vec{r}) \mathcal{W}[\hat{O}_2] (\vec{r}),
\end{equation}
in particular this can be used to express the Born rule
\begin{equation}\label{eq:wig_born_rule}
p(\vec{a}|\rho)= (2 \pi)^N \int_{\mathbb{R}^{2N}} \mathrm{d} \vec{r} \mathcal{W}[ \rho ] ( \vec{r}) \mathcal{W}[\Pi_{\vec{a}}] (\vec{r}),
\end{equation}
where $\int_\Omega \mathrm{d} \mu(\vec{a}) \Pi_{\vec{a}} = \id$ is a generic POVM.
The integral measure $\mu(\vec{a})$ on the outcome space $\Omega$ is generic, e.g. the the Lebesgue measure for general-dyne measurements or the counting measure for photon-counting measurements.
The Wigner functions of the effects of the POVM satisfy
\begin{equation}
\label{eq:wig_povm_identity}
\int_\Omega \mathrm{d} \mu(\vec{a}) \mathcal{W} \left[ \Pi_{\vec{a}} \right] (\vec{r}) = \frac{1}{(2 \pi)^N},
\end{equation}
where we used the linearity of the Wigner transform and the fact that $\mathcal{W}[\id](\vec{r}) = 1/ (2 \pi)^N$ (we assume that all the conditions to exchange the integration order hold true).
When the Wigner transform of the POVM effects $\mathcal{W} \left[ \Pi_{\vec{a}} \right] (\vec{r})$ is a Gaussian function we refer to these as Gaussian measurements (notice that we also consider the limiting case of homodyne detection, i.e. when the corresponding Wigner transform is a Dirac delta).

In this formalism we also recover the Wigner function of the partial trace as the marginal distribution~\cite{OzoriodeAlmeida2009}
\begin{equation}
\label{eq:wig_ptrace}
\mathcal{W} \left[ \Tr_B[\rho_{AB}] \right] (\vec{r}_A) = 	\int \mathrm{d} \vec{r}_B \mathcal{W}[\rho_{AB}] (\vec{r}_A,\vec{r}_B)
\end{equation}
and consequently the rule for the unnormalized conditional states after a measurement on one subsystem as
\begin{align}
\label{eq:wig_conditional_state}
&\mathcal{W} \left[ \Tr_B[\rho_{AB} \id \otimes \Pi_{\vec{a}} ] \right] (\vec{r}_A) = \nonumber \\
&= (2 \pi)^{N_B} \int \mathrm{d} \vec{r}_B \mathcal{W}[\rho_{AB}] (\vec{r}_A,\vec{r}_B) \mathcal{W}\left[ \Pi_{\vec{a}} \right](\vec{r}_B),
\end{align}
to get a normalized Wigner function the previous expression should be divided by the probability density $p(\vec{a})$ given by the Born rule~\eqref{eq:wig_born_rule}.
Moreover, the tensor product operation is simply given by multiplication of Wigner functions
\begin{equation}
\label{eq:wig_tensor_prod}
\mathcal{W}\left[ \hat{O}_1 \otimes \hat{O}_2 \right] (\vec{r}_1,\vec{r}_2) = \mathcal{W} \left[ \hat{O}_1 \right] (\vec{r}_1) \mathcal{W} \left[ \hat{O}_2 \right] ( \vec{r}_2).
\end{equation}

If the operator $\hat{O}$ is a density operator the Wigner transform gives the usual real valued Wigner function.
On the other hand for the operators $ | n \rangle \langle m| $ corresponding to the Fock basis $| n \rangle$ for a single mode, the Wigner function is not real but Hermitian, in the following sense
\begin{align}
\mathcal{W}[| n \rangle \langle m| ](x,p) & = \frac{1}{\pi} \int \mathrm{d}y \langle x + y |  n \rangle \langle m| x-y \rangle e^{-2 i y p} = \nonumber \\
& =  \frac{1}{\pi} \int \mathrm{d}y' \langle x - y' |  n \rangle \langle m| x+y' \rangle e^{2 i y' p} = \nonumber \\
& = W[| m \rangle \langle n| ]^*(x,p)
\end{align}

We conclude by noting that the Wigner transform is not the unique mapping between operators and phase space.
However, it is the only one with the following properties~\cite{Ferrie2011a,Bertrand1987}: it is real for any quantum state, it is a linear functional over density operators, its marginals are the probability distributions of canonical observables and the trace rule~\eqref{eq:tr_rule_ph_space} holds.

\subsection{Quadratic Hamiltonians and Gaussian states}\label{app:Gaussian}
We consider an Hamiltonian that is a quadratic polynomial in the canonical operators (the constant factors are irrelevant)
\begin{equation}
\label{eq:quad_Ham}
\hat{H}_G = \frac{1}{2} \hat{\vec{r}}^\mathsf{T} H \hat{\vec{r}} + \hat{\vec{r}}^\mathsf{T} \vec{h},
\end{equation}
where $H$ is a symmetric  $2 N \times 2 N $ matrix and $\vec{h}$ is a $2N$ real vector.
Such Hamiltonian corresponds to a linear transformation in Heisenberg picture (the so-called Gaussian unitaries)
\begin{equation}
\hat{U}_G^\dag \hat{\vec{r}} \hat{U}_G = S \hat{\vec{r}} + \vec{d},
\end{equation}
where $S$ is a symplectic matrix satisfying $S^\mathsf{T} \Omega S = \Omega$, which implies $\det S = 1$ (on a classical level this means that the phase space volume element is conserved).
More formally, the matrix $S$ belongs to the symplectic group ${\rm Sp}(2n,\mathbb{R})$, while the whole transformation belongs to the affine symplectic group ${\rm ISp}(2n,\mathbb{R})$.
A crucial property is that every Wigner function is covariant with respect to Gaussian unitary evolutions:
\begin{equation}
\label{eq:wigner_sympl}
\mathcal{W}[\hat{U}_G \rho \hat{U}_G^{\dag}] (\vec{r}) = \mathcal{W}[\rho] ( S^{-1} \vec{r} -  S^{-1}  \vec{d}).
\end{equation}

Gaussian states are defined as thermal states of Hamiltonians of the form~\eqref{eq:quad_Ham}, i.e.
\begin{equation}
\rho_G = \frac{e^{ -\beta \hat{H}_G } }{ \Tr \left[ e^{ -\beta \hat{H}_G }\right]},
\end{equation}
where pure states correspond to ground states and are obtained for $\beta \to \infty$.
A Gaussian state has a complete parametrization in term of first and second statistical moments (the covariance matrix $\vec{\sigma}$)
\begin{equation}
\vec{\bar{r}} = \Tr \left[ \rho_G \vec{\hat{r}} \right] \qquad \vec{\sigma} = \Tr \left[ \rho_G \left\{ (\vec{\hat{r}} - \vec{\bar{r}}), (\vec{\hat{r}} - \vec{\bar{r}})^\mathsf{T} \right\} \right]
\end{equation}
and its Wigner function is a Gaussian function of the form
\begin{equation}
\mathcal{W}\left[\rho_G\right] \left( \vec{r} \right) = \frac{1}{\pi^N \sqrt{\det \vec{\sigma}}} e^{- \left( \vec{r} - \vec{\bar{r}} \right)^\mathsf{T} \vec{\sigma}^{-1} \left( \vec{r} - \vec{\bar{r}} \right)} .
\end{equation}
For Gaussian states Eq.~\eqref{eq:wigner_sympl} entails a simple evolution in terms of covariance matrix and first moments:
\begin{equation}
\sigma'= S \sigma S^\mathsf{T} \qquad \bar{\vec{r}}'= S \bar{\vec{r}} + \vec{d} \; .
\end{equation}

Any covariance matrix can be diagonalized by symplectic transformations, each eigenvalue has multiplicity two and the $N$ different values $\nu_j$ are usually called symplectic eigenvalues.
The von Neumann entropy of a Gaussian state is a function of the symplectic eigenvalues only~\cite{serafini2017quantum,Holevo2001,Genoni2010} and one does not need the eigenvalues of the infinite dimensional density operator to compute it.
For a generic $N$-mode Gaussian state we have
\begin{equation}
S \left( \rho_G \right) = \sum_j^{N} h ( \nu_j ),
\end{equation}
where $\left\{ \nu_j  , j \in [1,\dots, N] \right\}$ are the symplectic eigenvalues and we defined the following function
\begin{equation}
h(x) = \left( \frac{x+1}{2} \right) \log \left( \frac{x+1}{2} \right) - \left( \frac{x-1}{2} \right) \log \left( \frac{x-1}{2} \right)  \;.
\end{equation}
In particular, for single mode systems the only symplectic eigenvalue is the square root of the determinant of the covariance matrix and we have
\begin{equation}
S( \rho_G ) = h( \sqrt{\det \sigma} ) \; .
\end{equation}

\subsection{Examples of Gaussian unitaries}\label{app:Gauss_unit}
An example of a Gaussian unitary is single mode squeezing
\begin{align}
\hat{S}(\xi) & = \exp \left[ \frac{\xi}{2}  \left( \hat{a}^\dag \right)^2  - \frac{\xi^*}{2} \hat{a}^2  \right] = \\
& = \exp \left[ \frac{i}{2} r \sin \psi \left( \hat{q}^2 - \hat{p}^2 \right) - \frac{i}{2} r \cos \psi \left( \hat{q}\hat{p} + \hat{p} \hat{q} \right) \right] 
\end{align}
where $\xi = r e^{i \psi}$, the corresponding symplectic matrix is thus 
\begin{equation}
S(r,\psi) = \begin{pmatrix} \cosh r  + \cos \psi \sinh r & \sin \psi \sinh r \\
\sin \psi \sinh r & \cosh r - \cos \psi \sinh r
 \end{pmatrix},
\end{equation}
for $\psi=0$ we get position squeezing, as in~\eqref{eq:squeezing_cubic_gate}.
Another important Gaussian unitary is the one for the beam splitter
\begin{align}
& \hat{U}_\mathsf{bs}(\phi,\theta) = \exp \left[ \phi e^{i \theta} \hat{a}_1^\dag \hat{a}_2 - \phi e^{-i \theta} \hat{a}_1 \hat{a}_2^\dag \right] = \\
& = \exp\left[ i \phi \cos \theta \left( \hat{p}_1 \hat{x}_2 - \hat{x}_1 \hat{p}_2 \right) + i \phi \sin \theta \left ( \hat{x}_1 \hat{x}_2 + \hat{p}_1 \hat{p}_2 \right) \right] \nonumber.
\end{align} 
In Section~\ref{sec:concentration} we parametrized it with the trasmissivity $T=\cos^2 \phi$ and we choose $\theta = \pi$; the corresponding symplectic matrix is then
\begin{equation}
S_\mathsf{bs}(T)= \begin{pmatrix} \sqrt{T} & 0 & \sqrt{1-T} & 0 \\ 
0 & \sqrt{T} & 0 & \sqrt{1-T} \\
 -\sqrt{1-T} & 0 & \sqrt{T} & 0  \\ 
0 & -\sqrt{1-T} & 0 & \sqrt{T}  \\ 
\end{pmatrix} \; .
\end{equation}


\section{Some parallels and connections with other resource theories}
\subsection{The maximal set of operations}
\label{app:maximalset}
Once a set of free state is chosen, one can define the maximal set of resource non-generating operations, i.e. the set of all CP trace non-increasing maps that map the set of free states into itself.
For this maximal set of free operations there are general results about conversion between resource states, both in the asymptotic~\cite{Brandao2015b} and single-shot regime~\cite{Gour2017}. Those proofs work only for discrete variables, but we intuitively expect the results to hold for CV as well (possibly by fixing the average energy of the states).

If we choose $\mathcal{G}$ as the set of free states, the corresponding resource non-generating operations is that of maximal Gaussian operations (MGO).
If instead we choose $\mathcal{W}_+$ we can call the resulting set maximal Wigner positive operations (MWPO).
In the resource theory of stabilizer quantum computation an explicit proof that the maximal set of free operations is indeed larger than stabilizer protocols has been recently given~\cite{Ahmadi2017}.
We conjecture that the same result should hold in the CV case; in our languange this means $\text{GP} \subset \text{MGO}$.
For the moment we do not have an explicit example to prove this conjecture.

An interesting comparison can be drawn with the resource theory of quantum coherence.
In~\cite{Chitambar2016b,Chitambar2016c} a critical examination of different kinds of free operations is made.
The physical set of incoherent operations can be used to distill a maximally coherent state in the many copies regime, however the process is strongly non-reversible and it is actually impossible to dilute a maximally coherent state into more weakly coherent states.
It would be interesting to answer similar questions for this resource theory, since many GPs can be implemented in the lab.

\subsection{Non-Gaussianity monotones and coherence monotones}

A different approach to introduce faithful monotones could be to connect the resource theory of quantum non-Gaussianity to the resource theory of coherence; something in this spirit has recently been proposed for the DV resource theory of magic~\cite{Mukhopadhyay2018}.
For CV, the resource theory of coherence has been generalized to the coherent state basis through a limiting procedure~\cite{Tan2017}; the resulting coherence monotone is also a monotone for the resource theory of linear optics, \textit{i.e.} passive Gaussian transformations.
This can be interpreted also as a resource theory of \emph{nonclassicality} for CV systems, where nonclassical states as those with a negative Glauber-Sudarshan $P$-function, instead of a negative Wigner function.
Remarkably, this resource theory has recently been studied in detail and linked to a metrological interpretation~\cite{Kwon2018,Yadin2018}.
We stress that this framework is entirely different from ours, since in our resource theory squeezing is assumed to be free.
However, for our purposes it might be possible to take similar steps and define a resource theory of superpositions of pure Gaussian states.
An option could be to generalize results from~\cite{Theurer2017}, where a general resource theory of superposition of arbitrary (\textit{i.e.}, non-orthogonal) states is presented, for finite dimensions.
We leave such questions open for future investigations.

\section{Proofs about monotones}\label{app:monotones}
\subsection{Wigner negativity}
\label{app:wigneg}
Here, we consider the monotone based the integral of the absolute value of the Wigner function, called Wigner negativity
\begin{equation}
\label{eq:WigNeg}
\mathcal{N}[\rho] = \int \mathrm{d} \vec{r} | W_\rho (\vec{r}) | - 1 \,.
\end{equation}
A part from trivially satisfying property 1 of Def.~\ref{def:monotone}, in the following we prove that it also satisfies all the properties implied by property 3.
The constant factor $-1$ in~\eqref{eq:WigNeg} does not affect the proofs, it is only needed to make the monotone take the value zero on states with a positive Wigner, so we just consider the integral in the following statements.

\begin{enumerate}
\item \emph{Invariance under Gaussian unitaries}\\
\begin{equation}
\mathcal{N} \left[ \rho \right] = \mathcal{ N} \left[ \hat{U}_G \rho \hat{U}_G^\dag \right]
\end{equation}
This directly follows from~\eqref{eq:wigner_sympl} by changing the integration variables, since $\det S = 1$.

\item \emph{Invariance under composition with Gaussian states}\\
\begin{equation}
\mathcal{N} \left[ \rho \otimes \rho_G \right] = \mathcal{N} \left[ \rho \right]
\end{equation}
This property directly follows from~\eqref{eq:wig_tensor_prod} and from the fact the Gaussian states have Gaussian (and thus positive) Wigner functions.

\item \emph{Non-increasing on average under Gaussian measurements}
\begin{equation}
\int \mathrm{d} \vec{\lambda} p \left( \vec{\lambda} \right) \mathcal{N}\left[ \rho_{\vec{\lambda}}\right] \leq \mathcal{N} \left[ \rho \right]
\end{equation}
We consider a Gaussian POVM $\int \mathrm{d} \vec{\lambda} \Pi_{\vec{\lambda}} = \id $, where $\mathcal{W}[\Pi_\lambda](\vec{r})$ are Gaussian functions.
Given an initial state $\rho$ we have the unnormalized post-measurement states $\sigma_{\vec{\lambda}} = \Tr_{B} \left[ \rho \Pi_{\vec{\lambda}} \right] $, the probability density $p(\vec{\lambda}) = \Tr \left[ \rho \Pi_{\vec{\lambda}} \right]$ and the normalized post measurement states $\rho_{\vec{\lambda}} = \sigma_{\vec{\lambda}} / p(\vec{\lambda})$.

Since the Wigner transform is a linear functional and probability density are positive we can write
\begin{equation}
\left| \mathcal{W}\left[ \int \mathrm{d} \vec{\lambda} p(\vec{\lambda})  \rho_{\vec{\lambda}} \right] (\vec{r}) \right| = \left| \int \mathrm{d} \vec{\lambda} p(\vec{\lambda})  \mathcal{W}\left[  \rho_\lambda \right] (\vec{r}) \right|, 
\end{equation}
and using linearity again we have that
\begin{equation}
\left| \int \mathrm{d} \vec{\lambda} p(\vec{\lambda})  \mathcal{W}\left[  \rho_{\vec{\lambda}} \right] (\vec{r}) \right| = 
\left| \int \mathrm{d} \vec{\lambda} \mathcal{W}\left[  \sigma_{\vec{\lambda}} \right] (\vec{r}) \right|,
\end{equation}
where $\sigma_{\vec{\lambda}}$ are unnormalized post measurement states, for which the Wigner function is given by~\eqref{eq:wig_conditional_state}.
We can thus write
\begin{align}
&\left| \int \mathrm{d} \vec{\lambda} \mathcal{W}\left[  \sigma_{\vec{\lambda}} \right] (\vec{r})\right| \leq  \int \mathrm{d} \vec{\lambda} \left| \mathcal{W}\left[  \sigma_{\vec{\lambda}} \right] (\vec{r}) \right| = \\
&= \int \mathrm{d} \vec{\lambda} \left| (2 \pi)^{N'} \int \mathrm{d} \vec{r}' \mathcal{W}[\rho] \left( \vec{r}, \vec{r}' \right) \mathcal{W}[\Pi_{\vec{\lambda}}] (\vec{r}') \right| \nonumber \\
&\leq \int \mathrm{d} \vec{\lambda}  (2 \pi)^{N'} \int \mathrm{d} \vec{r}' \left| \mathcal{W}[\rho] \left( \vec{r}, \vec{r}' \right) \mathcal{W}[\Pi_{\vec{\lambda}}] (\vec{r}') \right| = \nonumber \\
&= \int \mathrm{d} \vec{r}' \left| \mathcal{W}[\rho] \left( \vec{r}, \vec{r}' \right) \right|, \notag
\end{align}
where we used the integral triangle inequality and the POVM resolution of the identity for the Wigner functions~\eqref{eq:wig_povm_identity}.

Integrating both sides of this inequality w.r.t. $\vec{r}$ we get to the sought result:
\begin{equation}
\label{eq:neg_gauss_meas_ineq_chain}
\mathcal{N} \left[\int \mathrm{d} \vec{\lambda} p(\vec{\lambda})  \rho_{\vec{\lambda}} \right] \leq  \int \mathrm{d} \vec{\lambda}  p(\vec{\lambda}) \mathcal{N}\left[ \rho_{\vec{\lambda}} \right] \leq \mathcal{N}
 \left[ \rho \right],
 \end{equation}
where we also proved monotonicity under trace preserving operations.
By taking the appropriate limit these results hold also for homodyne measurements.

\item \emph{Non-increasing under partial trace}\\
\begin{equation}
\mathcal{N} \left[ \Tr_B \rho_{AB} \right] \leq \mathcal{N} \left[  \rho_{AB} \right]
\end{equation}
This property follows from~\eqref{eq:wig_ptrace} using again the integral triangle inequality as in the previous proof.

\item \emph{Convexity}\\
\begin{equation}
\mathcal{N} \left(  \int \mathrm{d} \nu p(\nu) \rho_{\nu} \right) \leq \int \mathrm{d} \nu p(\nu) \mathcal{N} \left( \rho_{\nu} \right)
\end{equation}
Trivially follows from the integral triangle inequality (that we have actually already used in the proof at point 3).

\end{enumerate}

To have an operational Gaussian protocol, i.e. in order to have non-zero conditional probabilities, we need to consider a finite region of the outcome space $\Omega_i$ and not a single outcome.
Formally this procedure amounts to a coarse-graining of the continuous outcomes into subsets, thus defining a new POVM.
The new effects for discrete outcomes are just obtained by integration $\Pi_i = \int_{\Omega_i} \mathrm{d} \vec{\lambda} \Pi_{\vec{\lambda}}$, where the outcome space is partitioned as $\Omega = \cup_i \Omega_i$.
The Wigner function of such POVMs is thus a statistical mixture of the original Wigner functions; as a consequence, since the original functions are always positive, the same is true for the coarse grained ones.
This implies that the previous proof can straight-forwardly extended to these coarse-grained POVMs.
The same inequality for coarse grained POVMs can be obtained by virtue of the convexity of $\mathcal{N}$.

Finally, we remark that the proofs discussed do not rely on the Gaussian character of states and POVM elements, but just on the positivity of their Wigner functions.

\subsection{Wigner logarithmic negativity}
\label{app:mana}
The WLN~\eqref{eq:CV_mana} can be defined in terms of the Wigner negativity~\eqref{eq:WigNeg} as 
\begin{equation}
\mathsf{W}[\rho] = \log \left( \mathcal{N}[\rho] + 1 \right).
\end{equation}
Properties 1, 2 and 4 follow directly from the fact that $\log (x + 1)$ is a monotonically increasing function of $x$.

For the WLN the chain of inequalities~\eqref{eq:neg_gauss_meas_ineq_chain} splits in two separate inequalities 
\begin{align}
\mathsf{W}\left[ \int \mathrm{d} \vec{\lambda} p (\vec{\lambda} )\rho_{\vec{\lambda}} \right] \leq \mathsf{W}[\rho] \label{eq:mana_ineq1}\\
\int \mathrm{d} \vec{\lambda} p (\vec{\lambda} ) \mathsf{W}\left[ \rho_{\vec{\lambda}} \right] \leq \mathsf{W}[\rho]. \label{eq:mana_ineq2}
\end{align}
The the first inequality represents monotonicity under deterministic operations and follows from~\eqref{eq:neg_gauss_meas_ineq_chain} and from the monotonicity of the logarithm.
Inequality~\eqref{eq:mana_ineq2} represents monotonicity on average and it is guaranteed by Jensen's inequality
\begin{equation}
\int \mathrm{d} \vec{\lambda} p(\vec{\lambda}) \log \left( \mathcal{N}\left[ \rho_{\vec{\lambda}} \right] + 1\right) \leq \log \left( 1 +  \int \mathrm{d} \vec{\lambda} p(\vec{\lambda}) \mathcal{N}\left[  \rho_{\vec{\lambda}} \right] \right),
\end{equation}
which holds since the logarithm is a concave function.
Notice that the same inequality is true also for the coarse grained case, since Jensen's inequality is valid both for discrete and continuous distributions.

Similarly to what happens in other resource theories, the logarithm makes this monotone additive for separable states $\mathsf{W}\left[ \rho_1 \otimes \rho_2 \right] = \mathsf{W}\left[ \rho_1 \right] + \mathsf{W} \left[ \rho_2 \right]$, but also breaks convexity.
To have a convex monotone after the logarithm, the original monotone should be log-convex, $f(p x + (1-p) y) \geq f(x)^p f(y)^{1-p}$, which is a stronger requirement than convexity.
The lack of convexity is the reason why inequalities~\eqref{eq:mana_ineq1} and~\eqref{eq:mana_ineq2} are not chained as~\eqref{eq:neg_gauss_meas_ineq_chain}.

\subsection{Uniqueness of negativity and logarithmic negativity}
The argument given in~\cite{Veitch2014} for the uniqueness of sum negativity can be adapted to CV, even though in this case we need to perform the limit to unphysical infinite energy states.
We present the single mode case for simplicity, the argument can be easily generalized to multi-mode states.

We assume to have two states $\rho$ and $\rho'$ with the same negativity (and thus also the same WLN) $\mathcal{N}\left[ \rho \right] = \mathcal{N}\left[ \rho' \right]$; we also assume that $\mathcal{M}_N$ is a generic monotone that depends only on the negative values of the Wigner function, but not on their position in phase space.
We define a function $f_\rho \left( x,y \right) = | W_\rho (x,y)| $ when $W_\rho (x,p)$ is negative and $f_\rho \left( x,y \right)=0$ everywhere else.
We note that $f_\rho \left( x,p \right) / \mathcal{N}\left[ \rho \right]$ is a well defined probability distribution.
If $\rho$ and $\rho'$ have negative values we build two ancillary states
\begin{equation}
\begin{split}
\sigma = \int \mathrm{d} x \mathrm{d} y f_\rho \left( x,y \right) | x \rangle \langle x | \otimes | y \rangle \langle y |, \\
\sigma' = \int \mathrm{d} x \mathrm{d} y f_{\rho'} \left( x,y \right) | x \rangle \langle x | \otimes | y \rangle \langle y |,
\end{split}
\end{equation}
where $|x \rangle$ are unnormalized position eigenstates, which are the infinite squeezing limit of Gaussian position squeezed states.
Therefore we have 
\begin{equation}
\label{eq:monotones_uniq}
\mathcal{M}_N \left[ \rho \right] = \mathcal{M}_N \left[ \rho \otimes \sigma' \right] = \mathcal{M}_N \left[ \rho' \otimes \sigma \right] = \mathcal{M}_N \left[ \rho' \right],
\end{equation}
the two outer equalities are due to the fact that $\sigma$ and $\sigma'$ are the limit of a sequence of states in $\mathcal{G}$.
The central inequality is due to the fact that the monotone $\mathcal{M}_N$ depends only on the negative value of the values of the Wigner function and not on their position. Note that the two Wigner functions have the same negative values, since the Wigner function of $\sigma$ and $\sigma'$ correspond exactly to the probability distribution function $f_\rho / \mathcal{N}[\rho]$ and $f_{\rho'}/\mathcal{N}[\rho']$, since the Wigner functions of $|x \rangle$ is a one dimensional $\delta$ function.
Equality~\eqref{eq:monotones_uniq} implies that $\mathcal{M}_N$ is a function of $\mathcal{N}$.
Suppose that $\mathcal{N}[\rho'] \geq \mathcal{N}[\rho]$, we can always find a new state $\sigma$ such that $\mathcal{M}_N [\rho'] = \mathcal{M}_N [\rho \otimes \sigma] \geq \mathcal{M}_N[\rho]$, where the last inequalities is due to the weak monotonicity property of monotones. This means that $\mathcal{M}_N$ is a monotonically increasing function of $\mathcal{N}$.

Furthermore, if we require that $\mathcal{M}_N$ is additive, i.e. $\mathcal{M}_N [\rho^{\otimes n} ] = n \mathcal{M}_N [\rho]$ we can also follow the proof given in~\cite{Zhu2017a} to show that $\mathcal{M}_N$ must correspond to the WLN $\mathsf{W}$ multiplied by an arbitrary positive constant.

\section{Preliminary numerical checks on the convex roof monotone} \label{app:numcheck_convex}

In order to check that property \textit{3.a} (monotonicity on average) holds for the monotone~\eqref{eq:convroof} defined from the convex roof of the relative entropy of non-Gaussianity we performed some preliminary numerical checks.

\begin{figure}[h!]
\includegraphics[width=.492\columnwidth]{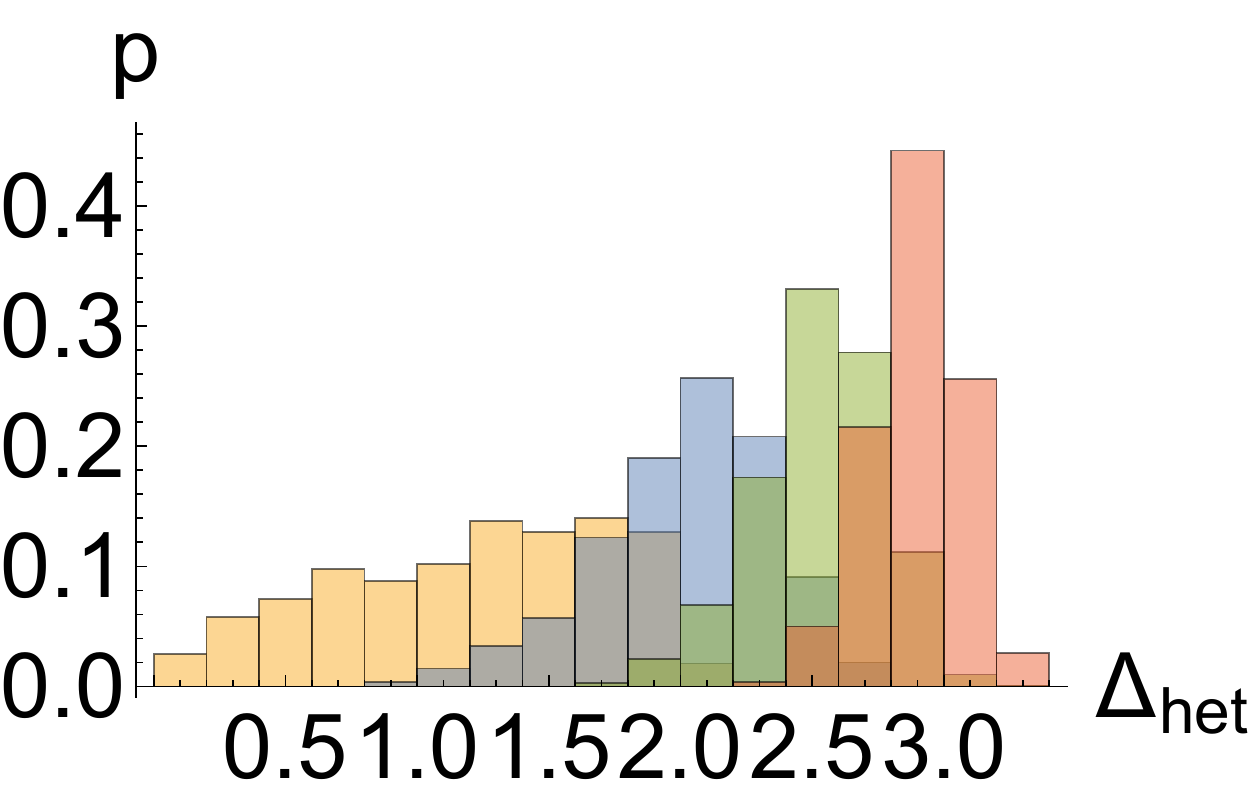}
\includegraphics[width=.492\columnwidth]{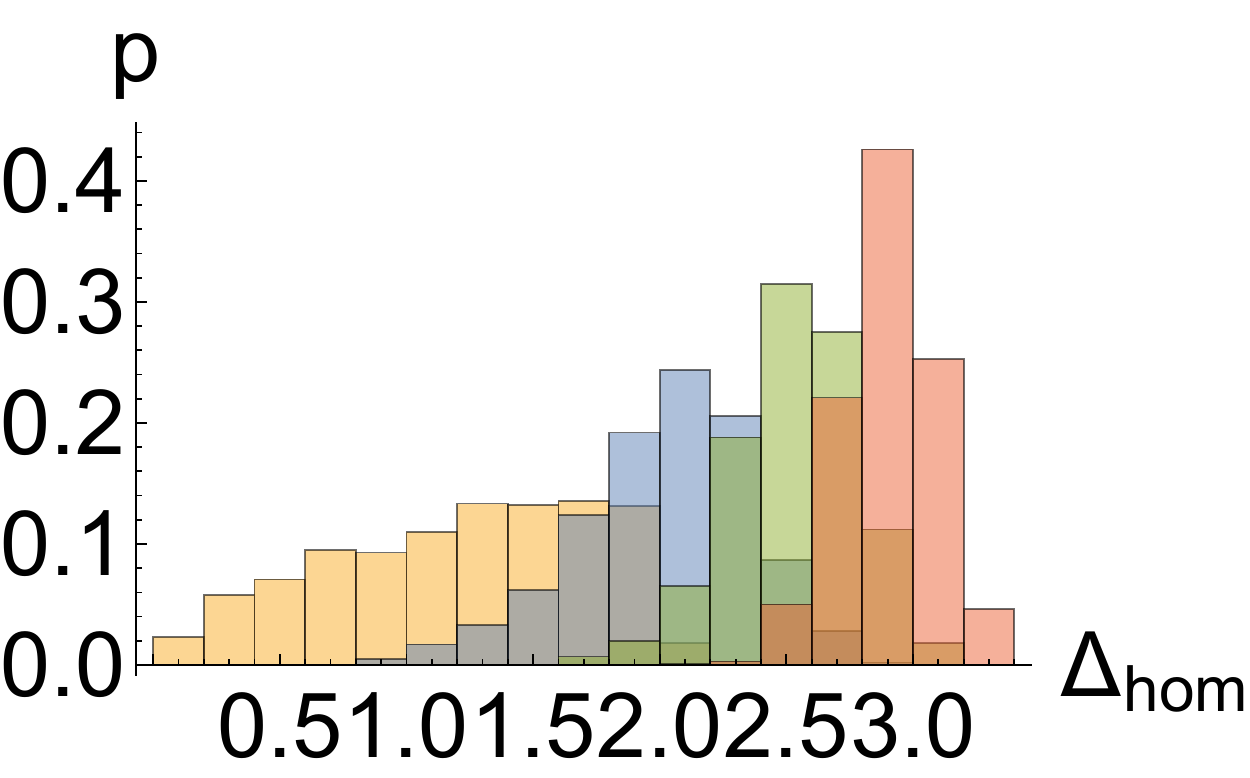}
\caption{Histograms of probabilities for the two quantities $\Delta_\mathsf{het}$ and $\Delta_\mathsf{hom}$.
We show data obtained by generating 1000 random states in bipartite Hilbert spaces of total dimension $N^2$ and local dimension $N$ with $N=2,3,4,5$ respectively corresponding to the orange, green, blue and red histograms (from left to right).
}\label{fig:cr_hist}
\end{figure}

We randomly generated states $| \Psi \rangle_{A,B}$ for two modes $A$ and $B$ with local Hilbert spaces of dimension $N$, with $N$ up to 5; we then performed heterodyne and homodyne measurements on the first mode and checked the validity of property \textit{3a}, i.e. that the following quantities are always positive
\begin{align}
& \Delta_\mathsf{het} = \delta \left[ | \Psi \rangle_{A,B} \right] - \int \frac{\mathrm{d}^2 \alpha}{\pi}  p \left( \alpha | \Psi  \right) \delta \left[ \frac{~_A\langle \alpha | \Psi \rangle_{A,B}}{\sqrt{p \left( \alpha | \Psi  \right)}} \right] \\
& \Delta_\mathsf{hom} = \delta \left[ | \Psi \rangle_{A,B} \right] - \int \mathrm{d} x \,  p \left( x | \Psi  \right) \delta \left[ \frac{~_A\langle x | \Psi \rangle_{A,B}}{\sqrt{p \left( x | \Psi  \right)}} \right] \;,
\end{align}
where $p( \alpha | \Psi ) = \sum_j  || \langle \alpha , j  | \Psi \rangle_{A,B} ||^2$, and $|\alpha,j\rangle = |\alpha\rangle_A \otimes |j\rangle_B$ (with $\{|j\rangle_B \}$ denoting the Fock basis for the Hilbert space of mode $B$) and analogously for the homodyne case.
An histogram showing this quantities is presented in Fig.~\ref{fig:cr_hist}.
Besides being always positive, the two differences $\Delta_\mathsf{het}$ and $\Delta_\mathsf{hom}$ also seem to be greater on average for random states generated in higher dimensional Hilbert spaces.

We could not find any violation of property \textit{3.a} for the monotone $\delta_\mathrm{CR}$, even though we need to keep in mind that the classes of states and the measurements considered in this numerical analysis are rather limited.

\bibliography{library}
\end{document}